\documentclass[journal]{IEEEtran}
\usepackage{blindtext}
\usepackage{graphicx}
\usepackage{cancel}
\usepackage[utf8]{inputenc}
\usepackage{graphics} 
\usepackage{epsfig} 
\usepackage{mathptmx} 
\usepackage{amsmath} 
\usepackage{amssymb}  
\usepackage{cite}
\usepackage{multicol}
\usepackage{mathtools, cuted}
\usepackage[cmintegrals]{newtxmath}
\usepackage{epstopdf}
\usepackage{graphics} 
\usepackage{epsfig} 

\usepackage{mathptmx} 
\usepackage{amsmath} 
\usepackage{amssymb}  
\usepackage{cite}
\usepackage{multicol}
\usepackage{mathtools, cuted}
\usepackage[cmintegrals]{newtxmath}
\usepackage{mathtools}

\usepackage{mathptmx}
\usepackage{helvet}
\usepackage{courier}
\usepackage{subfigure}
\usepackage{type1cm}
\usepackage{titlesec}
\usepackage{epsfig} 
\usepackage{mathptmx} 
\usepackage{amsmath} 
\usepackage{amssymb}  
\usepackage{cite}
\usepackage{multicol}
\usepackage{mathtools, cuted}
\usepackage[cmintegrals]{newtxmath}
\usepackage{makeidx}         
\usepackage{algorithmic}
\DeclareMathOperator*{\argmin}{\arg\!\min}
\DeclareMathOperator*{\argmax}{\arg\!\max}

 \usepackage{amsthm}

 \theoremstyle{definition}
\newtheorem{definition}{Definition}

\theoremstyle{theorem}
\newtheorem{theorem}{Theorem}

\theoremstyle{lemma}

\theoremstyle{remark}
\newtheorem{remark}{Remark}

 \theoremstyle{proposition}
\newtheorem{proposition}{Proposition}

 \theoremstyle{corollary}

 \theoremstyle{assumption}
\newtheorem{assumption}{Assumption}
\usepackage{multicol}        
\usepackage[bottom]{footmisc}
\usepackage{caption}
\usepackage{nomencl}
\usepackage[usenames, dvipsnames]{color}

\ifCLASSINFOpdf
\else
\fi
\begin{document}
%
\title{Safe Affine Transformation{\color{black}-Based Guidance} of a Large-Scale Multi-Quadcopter System (MQS)}
%
%
%

\author{Hossein Rastgoftar and Ilya Kolmanovsky
\thanks{The authors are with the Department
of Aerospace Engineering, University of Michigan, Ann Arbor,
MI, 48109 USA e-mail: {hosseinr,ilya}@umich.edu.}
}
%
%

\markboth{}%
{Shell \MakeLowercase{\textit{et al.}}: Bare Demo of IEEEtran.cls for IEEE Journals}
%



\maketitle
\vspace{-2cm}
\begin{abstract}
This paper studies the problem of affine transformation{\color{black}-based guidance} of a  multi-quadcopter system (MQS) {\color{black}in an obstacle-laden environment. {\color{black}Such MQSs can perform a variety of cooperative tasks including information collection, inspection
mapping, disinfection, and firefighting.} The MQS affine transformation is {\color{black}an approach to} a decentralized leader-follower coordination guided by $n+1$ leaders, where leaders are located at vertices of an $n$-D simplex, called \textit{leading simplex}, at any time $t$. The remaining agents are followers acquiring the desired affine transformation via local communication. {\color{black}Followers} are contained {\color{black}in} a rigid-size ball at any time $t$ but they can be distributed either inside or outside the leading simplex. By eigen-decomposition of the affine transformation coordination, safety in a large-scale MQS coordination can be ensured by constraining eigenvalues of the affine transformation. Given the initial and final configurations of the MQS, A* search is applied to optimally plan safe coordination of a large-scale MQS minimizing the travel distance between the the initial and final configuration.} The paper also proposes a proximity-based communication topology for followers to assign communication weights with their in-neighbors and acquire the desired coordination with minimal computation cost. 



\end{abstract}
\vspace{-0.5cm}
\section{Introduction}\label{Introduction}
Multi-agent coordination has been an active research area in the past few decades and found  various applications such as surveillance \cite{allouche2010multi}, {\color{black}search and rescue} missions \cite{kleiner2013rmasbench}, agriculture \cite{ali2010multi}, structural health monitoring \cite{yuan2005distributed}, and air traffic management \cite{idris2018air}. Consensus and containment control are common multi-agent coordination approaches that have been extensively studied in the past. 

Consensus control is the most well-known decentralized multi-agent coordination approach. Leaderless multi-agent consensus \cite{qin2016leaderless, ding2018leaderless} and leader-follower consensus \cite{wu2018leader} have {\color{black}been previously} proposed for multi-agent coordination applications. Multi-agent consensus under fixed communication topology and switching inter-agent communication  have been investigated in  \cite{wang2018fixed} and \cite{wen2016group}, respectively. Refs. \cite{zhou2018h, zhang2019delay} study stability of the consensus control in the presence of communication delays. Consensus control of a system of nonlinear agents {\color{black}has} {\color{black}been} investigated in Refs. \cite{du2020distributed, xu2016global}.

Containment control is a leader-follower method in which the group coordination is guided by a finite number of leaders and acquired by followers through local communication. Refs. \cite{cao2012distributed, ji2008containment} provide necessary and sufficient {\color{black}conditions} for stability and convergence {\color{black}in} the multi-agent {\color{black}containment} coordination problem. Containment under fixed and switching inter-agent communications are investigated in Refs. \cite{cao2012distributed, notarstefano2011containment, li2015containment} Also,  multi-agent containment control in the presence of time-varying delays are analyzed in \cite{shen2016containment, liu2014containment}. Refs. \cite{wang2013distributed, liu2015distributed}  have studied finite-time containment control of a multi-agent system.

Continuum deformation is another muti-agent coordination approach that treats agents as particles of a continuum{\color{black},} deforming in a $3$-D motion space. An $n$-D continuum deformation {\color{black}coordination} is guided by $n+1$ leaders in a $3$-D motion space where leaders are located at vertices of an $n$-D simplex at any time $t${\color{black},} and $n\in\{1,2,3\}$. In a continuum deformation coordination, desired trajectories {\color{black}are} planned by leaders {\color{black}and} acquired by followers through local communication \cite{rastgoftar2016continuum}. {\color{black}Therefore}, {\color{black}the} continuum deformation {\color{black}and containment control are both} decentralized leader-follower {\color{black}methods}. However, {\color{black}the} continuum deformation formally specifies and verifies safety in a large-scale agent coordination by ensuring inter-agent collision avoidance, obstacle collision avoidance, and agent containment \cite{rastgoftar2018safe, rastgoftar2019safe}. As the result, a large-scale multi-agent system can participate in a continuum deformation coordination mission and the agent team can aggressively deform to pass through the narrow passages in an obstacle-laden environments.

The existing continuum deformation approach \cite{rastgoftar2016continuum, rastgoftar2018safe} requires that the leaders form an $n$-D simplex at any time $t$. This requirement can be quite restrictive when agents are not uniformly distributed at the initial configuration. The main contribution of this paper is to advance the continuum deformation towards affine transformation in which  {\color{black}$n+1$ leaders defining the affine transformation coordination form} {\color{black}an} $m$-D polytope at any time $t${\color{black},} where $m\leq n\leq 3$. In other words, $n+1$ leaders, guiding {\color{black}the agent coordination,} are not required to form an $n$-D simplex at all times $t$. This advancement can significantly improves {\color{black}maneuverability} of a large-scale swarm coordination. {\color{black}In particular, our affine transformation-based coordination approach allows to plan more efficient motions than the existing continuum deformation approaches that it extends.}

This paper studies the problem of safe and scalable affine transformation of a multi-{\color{black}quadcopter} system (MQS) {\color{black}in an obstacle-laden environment {\color{black}(see Fig. \ref{example})}}. Compared to the existing literature and the authors' previous work, this paper offers the following novel contributions:
\begin{enumerate}
    \item{{\color{black}We decompose the affine transformation coordination problem into spatial and temporal planning problems. For the spatial planning, we use the A* search method to assign the optimal path of quadcopters such that the travel distance between their initial and final positions are minimized, and collision avoidance is guaranteed. For the temporal planning, the MQS travel time is determined such that deviation of every quadcopter from its global desired trajectory, defined by {\color{black}the} affine transformation, remains bounded at any time. } 
    }
    \item{This paper provides conditions {\color{black}guaranteeing} safety in a large-scale affine transformation coordination. By eigen-decomposition of the affine transformation {\color{black}and constraining the deformation eigenvalues of the affine transformation coordination},  inter-agent collision avoidance and quadcopter containment are ensured.}
    \item{This paper offers a new proximity-based communication topology for followers to acquire a desired affine transformation through local communication. {\color{black}Our approach is therefore of decentralized type.} 
    }
\end{enumerate}

{\color{black}The proposed affine transformation approach {\color{black}is particularly appealing {\color{black}for} application to} smart indoor or outdoor fire-fighting performed by a team of autonomous quadcopters {\color{black}exploiting the proposed approach}. In particular, a fire-fighter quadcopter team can effectively coordinate {\color{black}itself} in a geometrically-constrained and hazardous environment with minimal human interventions. The fire-fighter quadcopters can deform to pass through narrow channels and quicky react to a rapid growth of fire.}


This paper is organized as follows: Preliminaries {\color{black}of the} graph theory and motion space discretization are presented in Section \ref{Preliminary Notions and Definitions}. The problem of affine transformation {\color{black}coordination for} a large-scale MQS is stated in Section \ref{Problem Statement}. {\color{black}Section VI describes steps to determine ("tune") \cite{miranda2020parameter} algorithm parameters that we used in our case study. More specifically, affine transformation is defined  in section \ref{Problem 1: Affine Transformation Definition} and inferred via local communication in Section \ref{Problem 2: Affine Transformation Acquisition}}. Simulation results are presented in Section \ref{Simulation Results} and followed by concluding remarks in Section \ref{Conclusion}. {\color{black}The proofs are relegated to the Appendix \ref{Proofs}.  Quadcopter modeling details
are summarized in Appendices \ref{Quadcopter Angular Velocity and Acceleration} amd \ref{Time Derivatives of the Quadcopter Thrust Force}.}

%


\section{{\color{black}Preliminaries}}
\label{Preliminary Notions and Definitions}

\vspace{-.3cm}

\subsection{Graph Theory Notions}
\begin{figure}[ht]
\centering
\includegraphics[width=3.3 in]{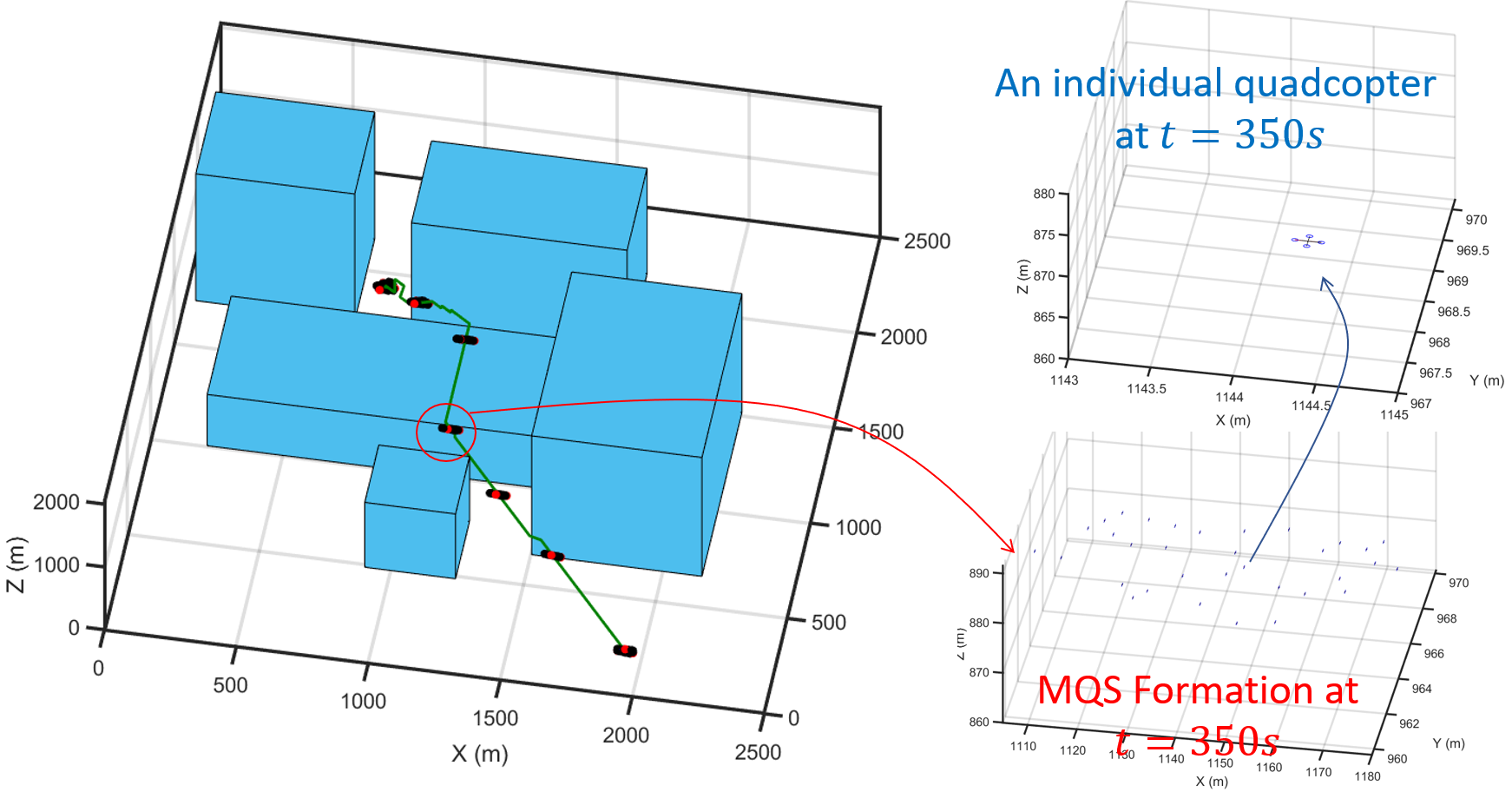}
\vspace{-.25cm}
\caption{Example MQS affine transformation coordination in an obstacle-laden motion space.}
\label{example}
\end{figure}

We consider an MQS consisting of $N$ {\color{black}quadcopters} moving collectively in a $3$-D {\color{black}space} where every quadcopter is uniquely identified by an index number $i\in \mathcal{V}=\{1,\cdots,N\}$ (see Fig. \ref{example}). By classifying quadcopters as leaders and followers, $\mathcal{V}$ can be expressed as $\mathcal{V}=\mathcal{V}_L\bigcup \mathcal{V}_F$, where $\mathcal{V}_L=\{1,\cdots,n+1\}$ and $\mathcal{V}_F=\mathcal{V}\setminus \mathcal{V}_L$ define the leaders' and followers' index numbers, respectively, in an $n$-D affine transformation, i.e. $n=1,2,3$.  While leaders move independently, followers acquire the desired coordination through local communication. Inter-agent communication is defined by digraph $\mathcal{G}\left(\mathcal{V},\mathcal{E}\right)$ with node set $\mathcal{V}$ and edge set $\mathcal{E}\subset \mathcal{V}\times \mathcal{V}$. 
Set $\mathcal{V}$ {\color{black}can} be expressed as $\mathcal{V}=\mathcal{V}_B\bigcup\mathcal{V}_I$ where $\mathcal{V}_B=\{1,\cdots,N_B\}$ and $\mathcal{V}_I\{N_B+1,\cdots,N\}$  define the index numbers of boundary and interior quadcopters, and $N_B\geq n+1$. Given edge set $\mathcal{E}$, {\color{black}the set of} in-neighbor quadcopters of quadcopter $i\in \mathcal{V}$ is defined by $\mathcal{N}_i=\{j\in \mathcal{V}\big|(j,i)\in \mathcal{E}\}$. 

\subsection{Position Notations}
This paper studies collective motion of $N$ quadcopters where {\color{black}the} position of every quadcopter is expressed with respect to an inertial coordinate system with base vectors $\hat{\mathbf{e}}_1=[1~0~0]^T$, $\hat{\mathbf{e}}_2=[0~1~0]^T$, and $\hat{\mathbf{e}}_3=[0~0~1]^T$. Throughout this paper, $\mathbf{r}_{i,0}=[x_{i,0}~y_{i,0}~z_{i,0}]^T$ and $\mathbf{r}_{i,f}=[x_{i,f}~y_{i,f}~z_{i,f}]^T$ denote the initial and final positions of quadcopter $i\in\mathcal{V}$ at {\color{black}the} initial time  $t_0$ and {\color{black}at the} final time $t_f$, respectively. Also, {\color{black}the} vector   $\mathbf{r}_i(t)=[x_{i}(t)~y_{i}(t)~z_{i}(t)]^T$ denotes the {\color{black}actual} position of {\color{black}quadcopter} $i$ at {\color{black}the} time {\color{black}instant} $t\in [t_0,t_f]$. {\color{black}The global} desired position of quadcopter $i\in \mathcal{V}$ is defined by an affine transformation as follows:
\begin{equation}
\label{1}
t\in [t_0,t_f],\qquad    \mathbf{r}_{i,a}(t)=\mathbf{Q}(t){\color{black}\left(\mathbf{r}_{i,0}-\mathbf{d}_0\right)}+\mathbf{d}\left(t\right),
\end{equation}
where $\mathbf{Q}(t)\in \mathbb{R}^{3\times 3}$ is the Jacobian matrix{\color{black},}  $\mathbf{d}(t)\in \mathbb{R}^{3\times 3}$ is the rigid-body displacement vector at time {\color{black}$t\in[t_0,t_f]$, and {\color{black}we let} $\mathbf{d}_0=\mathbf{d}\left(t_0\right)$}. Furthermore, 
\begin{equation}
\label{LocalDesiredEquation}
    \mathbf{r}_{i,d}(t)=
    \begin{cases}
    \mathbf{r}_{i,a}(t)&i\in \mathcal{V}_L\\
    \sum_{j\in \mathcal{N}_i}w_{i,j}\mathbf{r}_j{\color{black}(t)}&i\in \mathcal{V}_F
    \end{cases}
    .
\end{equation}
{\color{black}is called \textit{local desired position of quadcopter $i\in \mathcal{V}$} where $w_{i,j}>0$ is the communication {\color{black}weight} between follower $i$ and in-neighbor $j\in \mathcal{N}_i$ {\color{black}and $\mathbf{r}_j(t)$ is the actual position of quadcopter $j$}. Note that local and global desired positions of every leader $i\in \mathcal{V}_L$ are the same.}
 \begin{remark}
Elements of $\mathbf{Q}{\color{black}(t)}\in \mathbb{R}^{3\times 3}$ and $\mathbf{d}{\color{black}(t)=}\begin{bmatrix}d_1(t)&d_2(t)&d_3(t)\end{bmatrix}^T\in \mathbb{R}^{3\times 1}$ can be  uniquely related to the global desired positions of $n+1$ leader quadcopters where leader agents form an $n$-D simplex at initial time $t_0$ {\color{black}so that:}
\begin{equation}
\label{LeadersRank}
    \mathrm{rank}\left(
    \begin{bmatrix}
    \mathbf{r}_{2,0}-\mathbf{r}_{1,0}&\cdots&\mathbf{r}_{n+1,0}-\mathbf{r}_{1,0}
    \end{bmatrix}
    \right)=n.
\end{equation}
\end{remark}
\begin{assumption}\label{Assume1}
This paper assumes that quadcopters are initially distributed in {\color{black}an} $n$-D hyper-plane defined based on initial positions of leaders $1$ through $n+1$ guiding an $n$-D affine transformation. 
\end{assumption}
\begin{proposition}\label{th1}
If Assumption \ref{Assume1} is satisfied and leaders' initial positions satisfy rank condition \eqref{LeadersRank} at {\color{black}the} initial time $t_0$, initial position of every quadcopter $i\in \mathcal{V}$ can be uniquely expressed as a linear combination of leaders' initial positions {\color{black}so that}
\begin{equation}
\label{affineeeeelead}
    \forall i\in \mathcal{V},\qquad \mathbf{r}_{i,0}=\sum_{j=1}^{n+1}\alpha_{i,j}\mathbf{r}_{j,0},
\end{equation}
where
\begin{equation}
\label{alfasum}
\forall i\in \mathcal{V},\qquad \sum_{j=1}^{n+1}\alpha_{i,j}=1.
\end{equation}
\end{proposition}
\vspace{-.3cm}

\subsection{Rank Operator and Containment Function}
Let $\mathbf{p}_1$, $\mathbf{p}_2$, $\cdots$, $\mathbf{p}_{n+1}$ denote $n+1$ {\color{black}position vectors} in a $3$-D motion space. We define {\color{black}the} rank function {\color{black}as}
\begin{equation}
    \varrho_n\left(\mathbf{p}_1,\cdots,\mathbf{p}_{n+1}\right)={\color{black}\mathrm{rank}\left(
    \begin{bmatrix}
    \mathbf{p}_2-\mathbf{p}_1&\cdots&\mathbf{p}_{n+1}-\mathbf{p}_1
    \end{bmatrix}
    \right)}
    .
\end{equation}
Vectors $\mathbf{p}_1$, $\mathbf{p}_2$, $\cdots$, $\mathbf{p}_{n+1}$ define vertices of an $n$-D simplex, if $\varrho_n\left(\mathbf{p}_1,\cdots,\mathbf{p}_{n+1}\right)=n$. We also define {\color{black}the} containment function {\color{black}as}
\begin{equation}
    \varkappa_n\left(\mathbf{p}_1,\cdots,\mathbf{p}_n,\mathbf{c}\right)=\sum_{i=1}^{n+1}\mathrm{sign}\left(\mathbf{D}_i\right),
\end{equation}
where 
\begin{equation}
   \mathbf{D}_i=\begin{bmatrix}
    \mathbf{p}_1&\cdots&\mathbf{p}_{i-1}&\mathbf{c}&\mathbf{p}_{i+1}&\cdots&\mathbf{p}_{n+1}\\
    1&\cdots&1&1&1&\cdots&1\\
    \end{bmatrix}
\end{equation}
and $\mathbf{c}\in \mathbb{R}^{3\times 1}$ is the position of an arbitrary point in a $3$-D motion space, $\left|\mathbf{D}_i\right|$ is the determinant of matrix $\mathbf{D}_i$, and $\mathrm{sign}:\mathbb{R}\rightarrow\left\{-1,0,1\right\}$ is the sign function.

{\color{black}A} point $\mathbf{c}$ is inside an $n$-D simplex defined by $\mathbf{p}_1$, $\mathbf{p}_2$,$\cdots$, $\mathbf{p}_{n+1}$, if  $\left|\varkappa_n\left(\mathbf{p}_1,\cdots,\mathbf{p}_n,\mathbf{c}\right)\right|=n+1$ {\color{black}(See Ref. \cite{rastgoftar2018safe})}. Therefore, if $\varkappa_n\left(\mathbf{p}_1,\cdots,\mathbf{p}_n,\mathbf{c}\right)=n+1$ or  $\varkappa_n\left(\mathbf{p}_1,\cdots,\mathbf{p}_n,\mathbf{c}\right)=-\left(n+1\right)$, then, {\color{black}the} point $\mathbf{c}$ is inside the $n$-D simplex defined by $\mathbf{p}_1$ through $\mathbf{p}_{n+1}$. {\color{black}The rank} function $\varrho_n$ and {\color{black}the} containment function $\varkappa_n$ are used in Section \ref{Proximity-Based Inter-Agent Communication} to determine followers' in-neighbors and communication weights based on local proximity in the MQS initial configuration.

\vspace{-.3cm}

\subsection{Matrix Decomposition}
This paper uses {\color{black}the standard} $3-2-1$ Euler angles to define a rotation matrix by
\begin{equation}\label{rotationmatrix}
\resizebox{0.99\hsize}{!}{%
$
    \mathbf{R}\left({\color{black}X},{\color{black}Y},{\color{black}Z}\right)=\begin{bmatrix}
    C_{{\color{black}Y}} C_{{\color{black}Z}}&  C_{{\color{black}Y}} S_{{\color{black}Z}} &-S_{{\color{black}Y}}\\
S_{{\color{black}X}}S_{{\color{black}Y}} C_{{\color{black}Z}}-C_{{\color{black}X}}S_{{\color{black}Z}}&S_{{\color{black}X}}S_{{\color{black}Y}} S_{{\color{black}Z}}+C_{{\color{black}X}}C_{{\color{black}Z}}&S_{{\color{black}X}}C_{{\color{black}Y}}\\
   C_{{\color{black}X}}S_{{\color{black}Y}} C_{{\color{black}Z}}+S_{{\color{black}X}}S_{{\color{black}Z}} &C_{{\color{black}X}}S_{{\color{black}Y}} S_{{\color{black}Z}}-S_{{\color{black}X}}C_{{\color{black}Z}}&C_{{\color{black}X}}C_{{\color{black}Y}}
\end{bmatrix}
,
$
}
\end{equation}
where $C_{\left(\cdot\right)}$ and $S_{\left(\cdot\right)}$ abbreviate $\cos{\left(\cdot\right)}$ and $\sin{\left(\cdot\right)}$, respectively. Also, ${\color{black}X}$, ${\color{black}Y}$, and ${\color{black}Z}$ are the first, second, and third Euler angles where
\begin{equation}
    \mathbf{R}\left({\color{black}X},{\color{black}Y},{\color{black}Z}\right)=\mathbf{R}\left({\color{black}X},0,0\right)\mathbf{R}\left(0,{\color{black}Y},0\right)\mathbf{R}\left(0,0,{\color{black}Z}\right).
\end{equation}
Now, {\color{black}the} Jacobian matrix $\mathbf{Q}(t)$, introduced in Eq. \eqref{1}, can be {\color{black}represented} as follows:
{\color{black}
\begin{equation}
\label{PolarDecomposition}
    \mathbf{Q}(t)=\mathbf{\Phi}\left(\mathbf{\Theta}(t)\right),
\end{equation}
where
\begin{equation}
    \mathbf{\Theta}(t)=\begin{bmatrix}
    \lambda_1(t)&\cdots&\lambda_3(t)&\beta_1(t)&\cdots&\beta_6(t)
    \end{bmatrix}
    ^T
\end{equation}
is called {\color{black}the} \textit{deformation feature vector}, and $\mathbf{\Phi}$ {\color{black}can be decomposed as follows:}
\begin{equation}
    \mathbf{\Phi}=\mathbf{R}_r\mathbf{U}_D{\color{black},}
\end{equation}
{\color{black}where the matrix} $\mathbf{R}_r\left(\beta_1,\beta_2,\beta_3\right)$ is orthonormal{\color{black},} and {\color{black}the} {\color{black}deformation matrix} $\mathbf{U}_D\left(\beta_4,\beta_5,\beta_6,\lambda_1,\lambda_2,\lambda_3\right)$ is symmetric.
{\color{black}The} deformation features $\beta_1(t)$, $\beta_2(t)$, and $\beta_3(t)$ are the first, second, and third Euler angles, and 
\begin{equation}
    \mathbf{R}_r=\mathbf{R}\left(\beta_1(t),\beta_2(t),\beta_3(t)\right).
\end{equation}
{\color{black}The} deformation matrix {\color{black}can be represented as}
\begin{equation}
\label{UDecompose}
\mathbf{U}_D=\mathbf{R}_u\mathbf{\Lambda}\mathbf{R}_u^T{\color{black},}
 \end{equation}
where
 \begin{equation}
     \mathbf{R}_u=\mathbf{R}\left(\beta_4,\beta_5,\beta_6\right)
 \end{equation}
 is a rotation matrix, and $\beta_4$, $\beta_5$, and $\beta_6$ are the first, second, and third Euler angles. {\color{black}The matrix}
 \begin{equation}
     \mathbf{\Lambda}=\begin{bmatrix}
     \lambda_1&0&0\\
     0&\lambda_2&0\\
     0&0&\lambda_3
     \end{bmatrix}
 \end{equation}
 is diagonal and real.
 }
\begin{remark}
{\color{black}The} matrix $\mathbf{U}_D(t)$ can be expressed {\color{black}as}
\begin{equation}
\label{UDuseful}
    \mathbf{U}_D(t)=\sum_{i=1}^{3}\lambda_i(t)\hat{\mathbf{u}}_i\left(\beta_4(t),\beta_5(t),\beta_6(t)\right)\hat{\mathbf{u}}_i^T\left(\beta_4(t),\beta_5(t),\beta_6(t)\right),
\end{equation}
where 
\begin{equation}
\label{ui}
i=1,2,3,\qquad    \hat{\mathbf{u}}_i\left(\beta_4(t),\beta_5(t),\beta_6(t)\right)=\mathbf{R}^T\left(\beta_4(t),\beta_5(t),\beta_6(t)\right)\hat{\mathbf{e}}_i
\end{equation}
 is the $i$-th eigenvector of the deformation matrix $\mathbf{U}_D(t)$.
\end{remark}


  {\color{black}
\begin{proposition}
If $\lambda_1(t)=\lambda_2(t)=\lambda_3(t)=\lambda(t)$ at time $t$, then {\color{black}the} matrix $\mathbf{U}_D(t)$ simplifies {\color{black}to}
\begin{equation}
    \mathbf{U}_D(t)=\lambda(t)\mathbf{I}_3
\end{equation}\label{PROP2222}
{\color{black}independent of the values of} $\beta_4(t)$, $\beta_5(t)$, and $\beta_6(t)$ at $t$.
\end{proposition}
}
\vspace{-.3cm}
\section{Problem Statement {\color{black}and Solution Approach}}\label{Problem Statement}
This paper considers collective motion of a quadcopter team consisting of $N$ vehicles {\color{black}defined by $\mathcal{V}=\{1,\cdots,N\}$}, where quadcopter $i\in \mathcal{V}$ is modeled by
{\color{black}
\begin{equation}
\label{generalnonlineardynamics}
\begin{cases}
    \dot{\mathbf{x}}_i=\mathbf{f}_i\left(\mathbf{x}_i\right)+\mathbf{g}_i\left(\mathbf{x}_i\right)\mathbf{u}_i\\
    \mathbf{r}_i=\mathbf{C}_i\mathbf{x}_i
    \end{cases}
    .
\end{equation}
}
In \eqref{generalnonlineardynamics},
$
    \mathbf{x}_i=
    \begin{bmatrix}
    \mathbf{r}_i^T&\dot{\mathbf{r}}_i^T&\phi_i&\theta_i&\psi_i&{\color{black}\bf{\omega}}_i^T
    \end{bmatrix}
    ^T
$
is the state,  $\mathbf{u}_i=
    \begin{bmatrix}
    p_i&\tau_{\phi,i}&\tau_{\theta,i}&\tau_{\psi,i}
    \end{bmatrix}
    ^T$ is the input, {\color{black}$\mathbf{C}_i=\begin{bmatrix}
    \mathbf{I}_3&\mathbf{0}_{3\times 9}
    \end{bmatrix}$,}
   \[
    {\color{black}
   \resizebox{0.99\hsize}{!}{%
$
\mathbf{f}_i\left(\mathbf{x}_i\right)=
\begin{bmatrix}
\dot{\mathbf{r}}_i\\
{1\over m_i}p_i\hat{\mathbf{k}}_{b,i}-g\hat{\mathbf{e}}_3\\
\mathbf{\Gamma}_i^{-1}\left(\phi_i,\theta_i,\psi_i\right){\color{black}\bf{\omega}}_i\\
\mathbf{J}_i^{-1}{\color{black}\bf{\omega}}_i\times \left(\mathbf{J}_i{\color{black}\bf{\omega}}_i\right)\\
\end{bmatrix},
~\mathrm{and}~\mathbf{g}_i\left(\mathbf{x}_i\right)=
\begin{bmatrix}
\mathbf{0}_{3\times 1}&\mathbf{0}_{3\times 3}\\
{1\over m_i}\hat{\mathbf{k}}_{b,i}&\mathbf{0}_{3\times 3}\\
\mathbf{0}_{3\times 1}&\mathbf{0}_{3\times 1}\\
\mathbf{0}_{3\times 1}&\mathbf{J}_i^{-1}\\
\end{bmatrix}
,
$
}}
\]
where $m_i$ {\color{black}and $\mathbf{J}_i$ are}  the mass {\color{black}and mass moment of inertia} of quadcopter $i\in \mathcal{V}$, {\color{black}respectively,} $\mathbf{0}_{3\times 1}\in \mathbb{R}^{3\times 1}$, $\mathbf{0}_{3\times 3}\in \mathbb{R}^{3\times {\color{black}3}}${\color{black}, and $\mathbf{0}_{3\times 9}\in \mathbb{R}^{3\times {\color{black}9}}$} are the zero-entry matrices, $\mathbf{I}_3\in\mathbb{R}^{3\times 3}$ is the identity matrix, $g=9.81m/s^2$ is the gravity, $\mathbf{u}_i=\begin{bmatrix}
p_i&\tau_{\phi,i}&\tau_{\theta,i}&\tau_{\psi,i}
\end{bmatrix}
^T
$, and $\mathbf{\Gamma}_i\left(\phi_i,\theta_i,\psi_i\right)$ is defined in Eq. \eqref{Eq55} in Appendix {\color{black}\ref{Quadcopter Angular Velocity and Acceleration}}.

{\color{black}The} quadcopter team is positioned in an $n$-D hyper plane ($n=1,2,3$) at initial time $t_0$ {\color{black}and} the MQS initial formation is defined by set
    $\mathbf{\Omega}_0=\{\mathbf{r}_{1,0},\cdots\mathbf{r}_{N,0}\}$
at time $t_0$. It is desired that the MQS ultimately forms the desired configuration {\color{black}specified by}
    $\mathbf{\Omega}_f=\{\mathbf{r}_{1,f},\cdots\mathbf{r}_{N,f}\}$.
where
\begin{equation}
\label{finalposition}
\forall i\in \mathcal{V},\qquad    \mathbf{r}_{i,f}={\color{black}\bar{\mathbf{Q}}}_f\mathbf{r}_{i,0}+{\color{black}\bar{\mathbf{d}}}_f,
\end{equation}
{\color{black}and where the} matrix ${\color{black}\bar{\mathbf{Q}}_f}\in \mathbb{R}^{3\times 3}$ and vector ${\color{black}\bar{\mathbf{d}}}_f\in \mathbb{R}^{3\times 1}$ are known at time $t_f$, and the global desired trajectory of agent $i\in \mathcal{V}$ is defined by Eq. \eqref{1} for $t\in [t_0,t_f]$. {\color{black}To ensure safety, we {\color{black}require}}
that the MQS remains inside the {\color{black}rigid} containment ball
\begin{equation}
\label{St}
    \bar{\mathcal{S}}(t)=\left\{\mathbf{r}\in \mathbb{R}^3\big|\|\mathbf{r}-\left(\mathbf{d}(t)\right)\|\leq r_{\mathrm{max}}\right\}
\end{equation}
at any time $t\in [t_0,t_f]$. 

\begin{figure*}[ht]
\centering
\includegraphics[width=5.7 in]{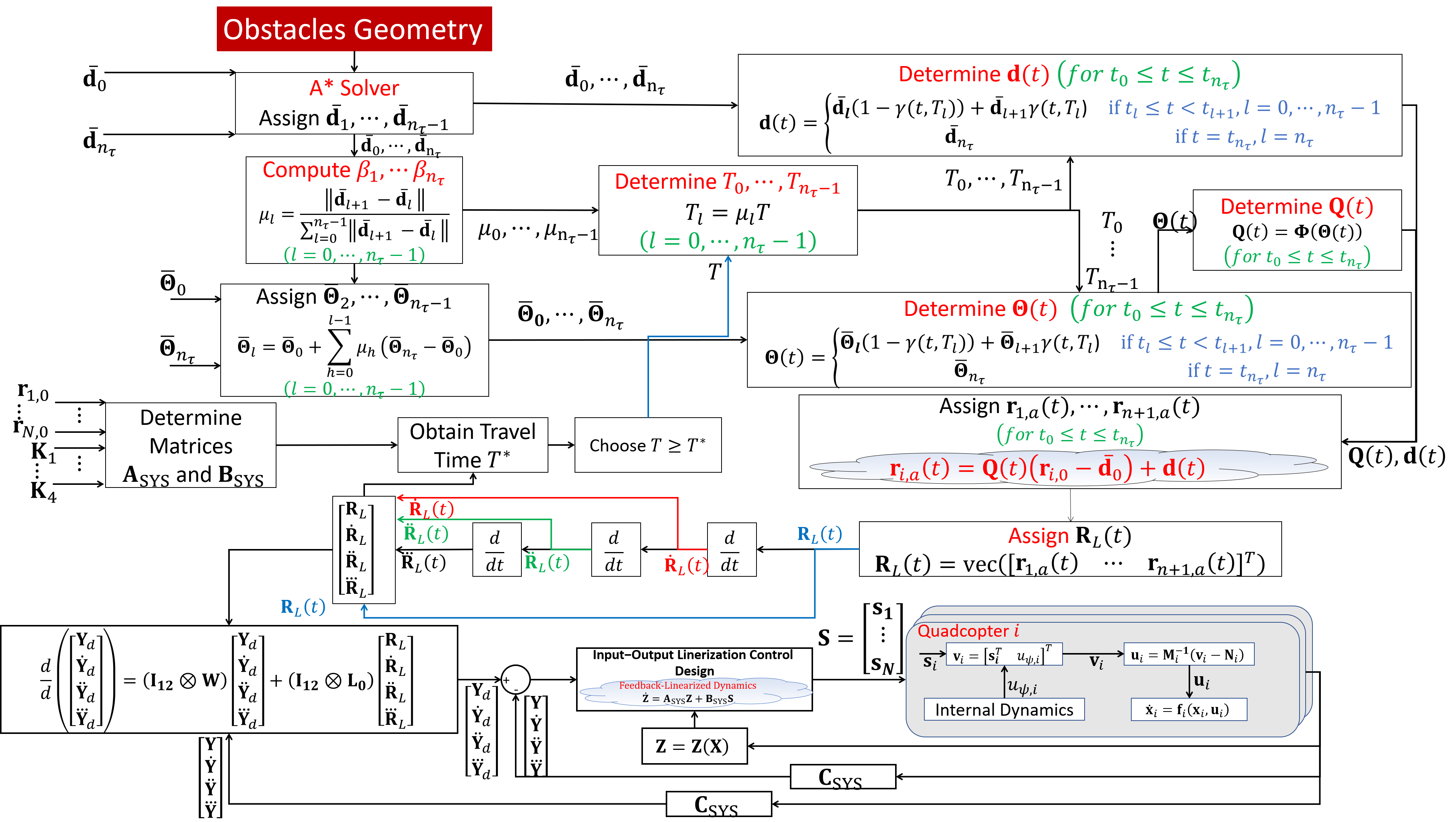}
\caption{Block diagram of the MQS collective dynamics {\color{black}with the proposed approach.}}
\label{Detecton2D}
\end{figure*}

Given the above problem setup, {\color{black}this paper offers a {\color{black}solution} shown in Fig. \ref{Detecton2D} to safely plan} affine transformation of a large-scale quadcopter team by {\color{black}addressing} the following two problems:

\textbf{Problem 1: Affine Transformation {\color{black}Determination}:} {\color{black}We {\color{black}determine} a safe MQS affine transformation by specifying {\color{black}the} Jacobian matrix $\mathbf{Q}(t)$ and {\color{black}the} rigid-body displacement vector $\mathbf{d}(t)$ such that {\color{black}the} travel distance between the initial and final configurations of the MQS is minimized in a geometrically-constrained motion space {\color{black}(see Fig. \ref{example})}. We assume that every quadcopter can be enclosed by a ball of radius $\epsilon$, and define {\color{black}the} matrix $\mathbf{Q}(t)$ over the time interval $ [t_0,t_f]$} such that no quadcopter collides {\color{black}with} obstacles, {\color{black}or with other quadcopters, and} {\color{black}followers do not leave the containment ball $\bar{\mathcal{S}}(t)$ {\color{black}defined by \eqref{St}} at any time $t\in [t_0,t_f]$.} 
{\color{black}Furthermore, we {\color{black}seek} to determine {\color{black}the} rigid-body displacement vector $\mathbf{d}$ minimizing the travel distance {\color{black}between the specified} initial and final conditions: $\bar{\mathbf{Q}}_0=\mathbf{Q}(t_0)$, $\bar{\mathbf{d}}_0=\mathbf{d}(t_0)$, $\bar{\mathbf{Q}}_f=\mathbf{Q}(t_f)$, $\bar{\mathbf{d}}_f=\mathbf{d}(t_f)$. 

}


\textbf{Problem 2: Affine Transformation Acquisition:} {\color{black}We} {\color{black}seek to} develop a decentralized method for acquiring the desired affine transformation {\color{black}with} local communication. In particular,  inter-agent communication and communication weights are assigned based on local proximity. Furthermore, {\color{black}we provide a condition {\color{black}guaranteeing} stability of the decentralized affine transformation coordination. We also {\color{black}seek to} ensure the boundedness of {\color{black}the} deviation of the quadcopter team from a desired affine tranformation coordination by choosing a sufficiently large travel time between in the initial and final MQS configurations.}
\section{Problem 1: Affine Transformation Definition}\label{Problem 1: Affine Transformation Definition}
{\color{black}An $n$-D affine transformation is defined by planning the {\color{black}trajectory of the} rigid-{\color{black}body} displacement vector $\mathbf{d}(t)$ and deformation vector $\mathbf{\Theta}(t)$ over the time-interval $t\in [t_0,t_f]$ as described in Sections \ref{Rigid-Body Displacement Vector} and \ref{Deformation Vector}}
\vspace{-.3cm}
\subsection{Planning of Rigid-Body Displacement Vector $\mathbf{d}$}\label{Rigid-Body Displacement Vector}
{\color{black}Given the initial and final displacement vectors $\bar{\mathbf{d}}_0$ and $\bar{\mathbf{d}}_f$, we use the the A* search to determine $n_\tau-1$ intermediate waypoints $\bar{\mathbf{d}}_1$, $\cdots$, $\bar{\mathbf{d}}_{n_\tau-1}$. The objective of the A* planner is to minimize the travel distance of the containment ball in an obstacle-laden motion space while ensuring collision avoidance.}

Given the optimal waypoints $\bar{\mathbf{d}}_0$, $\cdots$, $\bar{\mathbf{d}}_{n_\tau}$ ($\bar{\mathbf{d}}_{n_\tau}=\bar{\mathbf{d}}_f$), {\color{black}we define
\begin{equation}
    \mu_l=\dfrac{\|\bar{\mathbf{d}}_{l+1}-\bar{\mathbf{d}}_{l}\|}{\sum_{l=0}^{n_\tau-1}\|\bar{\mathbf{d}}_{l+1}-\bar{\mathbf{d}}_{l}\|}.
\end{equation}
for $l=0,1,\cdots,n_\tau-1$. {\color{black}
In this paper, 
\begin{equation}
\label{TL}
    T_l=\mu_l\left(t_{n_\tau}-t_0\right)
\end{equation}
is considered as the travel time between two consecutive waypoints $\bar{\mathbf{d}}_{l}$ and $\bar{\mathbf{d}}_{l+1}$, where $t_{n_\tau}=t_f$ is free.} 
}
{\color{black}The {\color{black}rigid body displacement
vector $\mathbf{d}(t)$ is defined by}
\begin{equation}
\mathbf{d}(t)=
\begin{cases}
    \bar{\mathbf{d}}_l\left(1-\gamma\left(t,T_l\right)\right)+\gamma\left(t,T_l\right)\bar{\mathbf{d}}_{l+1}&t_l\leq t<t_{l+1},~l<n_\tau\\
    \bar{\mathbf{d}}_{n_\tau}&t=n_\tau=t_f
\end{cases}
\end{equation}
where} $T_l=t_{l+1}-t_l$ is the travel time between $\bar{\mathbf{d}}_l$ and $\bar{\mathbf{d}}_{l+1}$,   $\bar{\mathbf{d}}_{n_\tau}=\mathbf{d}_f$, {\color{black}and $\gamma(t,T_l)$ is defined as follows:
\begin{equation}
 \label{poly1}
t\in [t_l,t_{l+1}],\qquad \gamma(t,T_l)=\sum_{j=0}^5\zeta_{j}\left({t-{\color{black}t_l}\over T_l}\right)^j
\end{equation}
{\color{black}for $l=0,\cdots,n_{\tau-1}$. Here} $\zeta_0$ through $\zeta_5$ are constant coefficients{\color{black},} and $T_l=t_{l+1}-t_l$. Therefore, the containment ball moves on a straight path at every time $t\in [t_l,t_{l+1}]$ 
where $\gamma(t_l,T_l)=0$ and $\gamma(t_{l+1},T_l)=1$ for every $T_l>0$ and $l=0,\cdots,n_\tau-1$. }

\vspace{-.3cm}

\subsection{Planning of Deformation Feature Vector Trajectory $\mathbf{\Theta(t)}$}\label{Deformation Vector}
{\color{black}By using Eq. \eqref{PolarDecomposition}, $\mathbf{Q}(t)$ can be {\color{black}expressed as $\mathbf{Q}(t)=\mathbf{\Phi}\left(\mathbf{\Theta}(t)\right)$, and assigned} by planning of the deformation vector $\mathbf{\Theta}(t)$ over the time-interval $[t_0,t_f]$.
{\color{black}The} deformation feature vector $\mathbf{\Theta}:\left[t_0,t_f\right]\rightarrow \mathbb{R}^{6\times 1}$ is defined as follows:
\begin{equation}
\label{ThetaDefinition}
\mathbf{\Theta}(t)=
\begin{cases}
    \bar{\mathbf{\Theta}}_l\left(1-\gamma\left(t,T_l\right)\right)+\gamma\left(t,T_l\right)\bar{\mathbf{\Theta}}_{l+1}&t_l\leq t<t_{l+1},~l<n_\tau\\
    \bar{\mathbf{\Theta}}_{n_\tau}&t=n_\tau=t_f
\end{cases}
,
\end{equation}
where initial and final conditions
\begin{subequations}
\begin{equation}
\bar{\mathbf{\Theta}}_0=\mathbf{\Theta}\left(t_0\right)=\begin{bmatrix}
   \lambda_{1,0}&\lambda_{2,0}&\lambda_{3,0}&\beta_{1,0}&\cdots&{\color{black}\beta_{6,0}}
    \end{bmatrix}
    ^T,
\end{equation}
\begin{equation}
\bar{ \mathbf{\Theta}}_f=\mathbf{\Theta}\left(t_f\right)=\begin{bmatrix}
   \lambda_{1,f}&\lambda_{2,f}&\lambda_{3,f}&\beta_{1,f}&\cdots&\beta_{6,f}
    \end{bmatrix}
    ^T
\end{equation}
\end{subequations}
are known, and
\begin{equation}
    \bar{\mathbf{\Theta}}_l=\bar{\mathbf{\Theta}}_0+\mu_l\left(\bar{\mathbf{\Theta}}_f-\bar{\mathbf{\Theta}}_0\right).
\end{equation}
Note that function {\color{black}$\gamma(t,T_l)$} is defined in Eq.  \eqref{poly1}.}

\subsubsection{Deformation Angles $\beta_4$, $\beta_5$ and $\beta_6$}
{\color{black}
There is no constraint on selecting  $\beta_{4,0}$, $\beta_{5,0}$, $\beta_{6,0}$, {\color{black}and they} can be arbitrarily because $\lambda_{1,0}=\lambda_{2,0}=\lambda_{3,0}=1$.} In this paper, we {\color{black}let shear deformation angles $\beta_5(t)$ and $\beta_6(t)$ be constant {\color{black}over time}, and assign} them based on {\color{black}the} initial positions of the quadcopters 
by solving the following max-min optimization problem:
\begin{equation}
\label{beta5beta6}
    \left(\beta_{5,0},\beta_{6,0}\right)=\argmax\limits_{\beta_5,\beta_6}\left\{\min\limits_{i,j\in \mathcal{V},~i\neq j}\left\{\left(\mathbf{r}_{i,0}-\mathbf{r}_{j,0}\right)\cdot \hat{\mathbf{u}}_1\left(0,\beta_5,\beta_6\right)\right\}\right\}.
\end{equation}
Note that $\hat{\mathbf{u}}_{1,0}$ is independent of $\beta_{4,0}$ {\color{black}and hence} we choose $\beta_{4,0}=0$ without loss of generality. Therefore,
\begin{subequations}
\begin{equation}
    \hat{\mathbf{u}}_{1,0}=
\begin{bmatrix}
\cos\beta_{5,0}\cos\beta_{6,0}&\cos\beta_{5,0}\sin\beta_{6,0}&-\sin\beta_{5,0}
\end{bmatrix}
^T,
\end{equation}
\begin{equation}
    \hat{\mathbf{u}}_{2,0}=
\begin{bmatrix}
\sin\beta_{6,0}&\cos\beta_{6,0}&0
\end{bmatrix}
^T,
\end{equation}
\begin{equation}
    \hat{\mathbf{u}}_{3,0}=
\begin{bmatrix}
\sin\beta_{5,0}\cos\beta_{6,0}&\sin\beta_{5,0}\sin\beta_{6,0}&\cos\beta_{5,0}
\end{bmatrix}
^T
\end{equation}
\end{subequations}
are the eigenvectors of deformation matrix {\color{black}$\mathbf{U}_D(t)$ at any time $t\in [t_0,t_f]$}.

{\color{black}
\begin{remark}
Deformation angles $\beta_{5,0}$ and $\beta_{6,0}$ are assigned such that the unit vector $\hat{\mathbf{u}}_{1,0}$ {\color{black}is} along the line connecting the two {\color{black}quadcopters} identified by solving \eqref{beta5beta6}.
\end{remark}
}

\subsubsection{Eigenvalues $\lambda_1$, 
$\lambda_2$, and $\lambda_3$}
{\color{black}Theorem \ref{thmlambdamax} is provided in this section to ensure inter-agent collision avoidance and quadcopter containment by assigning lower and upper bounds on eigenvalues $\lambda_1$, $\lambda_3$, and $\lambda_3$.}

 \begin{definition}
{\color{black}The minimum global separation distance in an affine transformation is defined by}
\begin{equation}
\label{dminnnnnnn}
    d_{\mathrm{min}}=\min\limits_{i,j\in \mathcal{V},~i\neq j}\left\{\left(\mathbf{r}_{i,0}-\mathbf{r}_{j,0}\right)\cdot \hat{\mathbf{u}}_1\left(0,\beta_5,\beta_6\right)\right\}.
\end{equation}
\end{definition}
{\color{black}
\begin{definition}
{\color{black}The maximum global separation distance in an affine transformation is defined as}
\begin{equation}
\label{dmaxxx}
    d_{\mathrm{max}}=\max\limits_{i\in \mathcal{V}}=\|\mathbf{r}_{i,0}-\bar{\mathbf{d}}_0\|_2.
\end{equation}
\end{definition}
 }
 \begin{theorem}\label{thmlambdamax}
Assume every quadcopter is enclosed by a ball of radius $\epsilon$, and the control input $\mathbf{u}_i$ is designed such that
\begin{equation}
\label{eqreferror12}
\|\mathbf{r}_i(t)-\mathbf{r}_{i,a}(t)\|\leq \delta
\end{equation}
at any time {\color{black}$t\in[t_0,t_f]$} where $\delta>0$ is constant. {\color{black}Inter-agent collision avoidance and quadcopter containment conditions, specified by 
\begin{subequations}
\begin{equation}
\forall t\in \left[t_0,t_f\right],\qquad     \bigwedge_{i=1}^{N-1}\bigwedge_{j=i+1}^N\left(\|\mathbf{r}_i(t)-\mathbf{r}_j(t)\|> 2\epsilon\right),
\end{equation}
\begin{equation}
\forall t\in \left[t_0,t_f\right],\qquad     \bigwedge_{i=1}^{N}\left(\mathbf{r}_i(t)\in \bar{\mathcal{S}}(t)\right)
\end{equation}
\end{subequations}
 are} guaranteed, if 
{\color{black}
\begin{subequations}
\begin{equation}
\label{intagcol}
\forall t\in [t_0,t_f],\qquad   \lambda_1(t)> \lambda_{\mathrm{min}},
\end{equation}
\begin{equation}
\label{quadcontainmentttttt}
\forall t\in [t_0,t_f],~i=1,2,3,\qquad   \left|\lambda_i(t)\right|\leq \lambda_{\mathrm{max}},
\end{equation}
\end{subequations}
 where}
\begin{subequations}
\begin{equation}
\label{lmin}
    \lambda_{\mathrm{min}}=\dfrac{2\left(\delta+\epsilon\right)}{d_{\mathrm{min}}},
\end{equation}
{\color{black}
\begin{equation}
\label{lmax}
    \lambda_{\mathrm{max}}=\dfrac{r_{\mathrm{max}}-\delta-\epsilon}{d_{\mathrm{max}}}.
\end{equation}
}
\end{subequations}
\end{theorem}
{\color{black}
Note that inter-agent collision avoidance is guaranteed only by imposing constraint \eqref{intagcol} on eigenvalue $\lambda_1$. However, all eigenvalues of matrix $\mathbf{U}_D(t)$ must satisfy  safety condition \eqref{quadcontainmentttttt} to ensure no quadcopter leaves the containment ball $\bar{\mathcal{S}}(t)$ at any time $t\in [t_0,t_f]$.

\begin{remark}
While $\lambda_{1,0}=\lambda_{2,0}=\lambda_{3,0}=1$, $\lambda_{1,f}$, $\lambda_{2,f}$, and $\lambda_{3,f}$ need to be selected such that Eq. \eqref{finalposition} is satisfied, and
\begin{equation}
    i=1,2,3,\qquad \left|\lambda_{i,f}\right|\leq \lambda_{\mathrm{max}}.
\end{equation}
\end{remark}
\subsubsection{{\color{black}Rotation angles $\beta_1$, $\beta_2$, and $\beta_3$}}
The  initial values of rotation angles $\beta_1$, $\beta_2$, and $\beta_3$, denoted by $\beta_{1,0}$, $\beta_{2,0}$, $\beta_{3,0}$, can be arbitrarily selected. However, $\beta_{1,f}$, $\beta_{2,f}$, $\beta_{3,f}$ need to be selected such that Eq. \eqref{finalposition} is satisfied.
}


\vspace{-.3cm}


\section{Problem 2: Affine Transformation Acquisition}\label{Problem 2: Affine Transformation Acquisition}
{\color{black}In this paper, a} desired affine transformation is acquired in a decentralized fashion via local communication. {\color{black}A} proximity-based communication topology is developed in Section  \ref{Proximity-Based Inter-Agent Communication} to (i) classify  {\color{black}quadcopters} into followers and leaders and (ii) determine in-neighbor  {\color{black}quadcopters} of every follower  {\color{black}quadcopter} $i\in \mathcal{V}_F$. MQS collective dynamics {\color{black}are} then obtained in Section \ref{MAS Collective Dynamics1} {\color{black}and followed by} analysis of stability and boundedness of the MQS collective dynamics in {\color{black}Sections \ref{MAS Collective Dynamics2} and \ref{Inter-Agent Cion Avoidance}, respectively.}

\subsection{Proximity-Based Inter-Agent Communication}
\label{Proximity-Based Inter-Agent Communication}
In {\color{black}a decentralized affine transformation coordination}, leaders move independently therefore $\mathcal{N}_i=\emptyset$ if $i\in \mathcal{V}_L$. Non-leader boundary {\color{black}quadcopters, defined by $\mathcal{V}_B$,} directly communicate with leaders. Therefore,
\begin{equation}
    i\in \mathcal{V}_B\setminus\mathcal{V}_L,\qquad \mathcal{N}_i=\mathcal{V}_L.
\end{equation}

In-neighbors of the interior {\color{black}quadcopters, defined by $\mathcal{V}_I$,} are assigned based on local proximity. For every {\color{black}quadcopter} $i\in \mathcal{V}_I$, {\color{black}we define} $l_i$-proximity set 
\begin{equation}
\begin{split}
    \mathcal{W}_{i,n}\left(l_i\right)=&\bigg\{\left(i_1,\cdots,i_{n+1}\right)\in \mathcal{V}^{n+1}\big|\left(\bigwedge_{k=1}^{n+1}\|\mathbf{r}_{i_k,0}-\mathbf{r}_{i,0}\| \leq l_i\right)\wedge\\
    &\left(\left|\varkappa_n\left(\mathbf{r}_{i_1,0},\cdots,\mathbf{r}_{i_{n+1},0},\mathbf{r}_{i,0}\right)\right|=n+1\right)\bigg\}
\end{split}
\end{equation}
{\color{black}as the set of all} $n$-D simplexes that are inside {\color{black}the} ball of radius $l_i$ with the center positioned at $\mathbf{r}_{i,0}$. Then, the minimum proximity distance $l_i^*$ is assigned by solving the following optimization problem:
\begin{equation}
\label{opt33}
    \min l_i
\end{equation}
such that
\begin{equation}
\label{opt34}
    \mathcal{W}_i\left(l_i\right)\neq \emptyset.
\end{equation}
\begin{remark}
If $\left|\mathcal{W}_i\left(l_i^*\right)\right|=1$, then, $\mathcal{W}_i=\{\mathcal{N}_i\}$. 
\end{remark}
\begin{assumption}
If $\left|\mathcal{W}_i\left(l_i^*\right)\right|>1$, {\color{black}any collection of $n+1$ quadcopters} belonging to set $\mathcal{W}_i\left(l_i^*\right)$ can be selected as in-neighbors of interior agent $i\in \mathcal{V}_I$.
\end{assumption}

{\color{black}Let the} communication {\color{black}weight} of follower {\color{black}quadcopter} $i\in \mathcal{V}_F$ with in-neighbor {\color{black}quadcopter} $j\in \mathcal{N}_i$ is denoted by $w_{i,j}$. Let $N_{i}=\{i_1,\cdots,i_{n+1}\}$ define in-neighbors of follower $i\in \mathcal{V}_F$, then,  the local desired {\color{black}trajectory} of quadcopter $i\in \mathcal{V}_F$ is given by 
\begin{equation}
    \forall t\in [t_0,t_f],i\in \mathcal{V}_F,\qquad \mathbf{r}_{i,d}(t)=\sum_{k=1}^{n+1}w_{i,k}\mathbf{r}_{i_k}(t),
\end{equation}
where $\mathbf{r}_{i_k}(t)$ denotes the actual position of in-neighbor $i_k\in \mathcal{N}_i$ ($k=1,\cdots,n+1$). Communication weights of follower $i\in \mathcal{V}_F$ are {\color{black}defined as} {\color{black}\cite{rastgoftar2016continuum, rastgoftar2018safe}}
\begin{subequations}
\label{comweightss}
\begin{equation}
    \begin{bmatrix}
    w_{i,i_2}\\
    \vdots\\
    w_{i,i_{n+1}}
    \end{bmatrix}
    =\begin{bmatrix}
    \mathbf{r}_{i_2,0}-\mathbf{r}_{i_1,0}&\cdots&\mathbf{r}_{i_{n+1},0}-\mathbf{r}_{i_1,0}
    \end{bmatrix}
    ^{-1}\mathbf{r}_{i,0},
\end{equation}
\begin{equation}
    w_{i,i_1}=1-\sum_{k=2}^{n+1}w_{i,i_k}.
\end{equation}
\end{subequations}

Given followers' communication weights, weight matrix $\mathbf{W}=[W_{ij}]\in \mathbb{R}^{N\times N}$ is defined as follows:
\begin{equation}
    W_{ij}=
    \begin{cases}
    w_{i,j}&i\in\mathcal{V}_F,~j\in \mathcal{N}_i\\
    0&\mathrm{otherwise}
    \end{cases}
    .
\end{equation}
{\color{black}The} matrix $\mathbf{W}$ can be partitioned as follows:
\begin{equation}
\label{WW}
    \mathbf{W}=
    \left[
    \begin{array}{c|c}
    \mathbf{0}&\mathbf{0}\\
    \hline
    \mathbf{F}&\mathbf{G}
    \end{array}
    \right]\in \mathbb{R}^{N\times N},
\end{equation}
where $\mathbf{F}\in \mathbb{R}^{\left(N-n-1\right)\times \left(n+1\right)}$; matrix $\mathbf{G}\in \mathbb{R}^{\left(N-n-1\right)\times \left(N-n-1\right)}$ is non-negative.

\begin{theorem}\label{th2}
Assume inter-agent communication is defined by graph $\mathcal{G}\left(\mathcal{V},\mathcal{E}\right)$ with node set $\mathcal{V}$ and edge set $\mathcal{E}\subset \mathcal{V}\times \mathcal{V}$, where $\mathcal{V}=\mathcal{V}_B\bigcup\mathcal{V}_I$; $\mathcal{V}_B=\{1,\cdots,N_B\}$ and $\mathcal{V}_I=\{N_B+1,\cdots,N\}$ define index numbers of boundary and interior agents, respectively. If leaders defined by set $\mathcal{V}_L=\{1,\cdots,n+1\}\subset \mathcal{V}_B$ moves independently, non-leader boundary agents defined by $\mathbf{V}_B\setminus \mathcal{V}_L$ all communicate with leaders, followers' in-neighbors are determined by solving the optimization problem given in \eqref{opt33} and \eqref{opt34}, and followers communication weights are defined based on agents' initial positions using relation \eqref{comweightss}, then, {\color{black}the} matrix 
\begin{equation}
    \mathbf{L}=-\mathbf{I}+\mathbf{W}
\end{equation}
is Hurwitz.
\end{theorem}


Let
\begin{equation}
\label{Ya}
    \mathbf{Y}_a(t)=\mathrm{vec}\left(
    \begin{bmatrix}
    \mathbf{r}_{1,a}(t)&\cdots&\mathbf{r}_{N,a}(t)
    \end{bmatrix}^T\right)
\end{equation}
{\color{black}aggregate} global desired positions of all {\color{black}quadcopters} at time $t$ and 
\begin{equation}
\label{RL}
    \mathbf{R}_L(t)=\mathrm{vec}\left(
    \begin{bmatrix}
    \mathbf{r}_{1,a}(t)&\cdots&\mathbf{r}_{n+1,a}(t)
    \end{bmatrix}^T\right)
\end{equation}
{\color{black}aggregate} global desired positions of all leaders at time $t${\color{black}, where $\mathrm{vec}\left(\cdot\right)$ is the matrix vectorization operator.} Vectors $\mathbf{Y}_a(t)$ and $\mathbf{R}_L(t)$ are related by \cite{rastgoftar2018safe}
\begin{equation}
    \mathbf{Y}_a(t)=\left(\mathbf{I}_3\otimes \mathbf{H}\right)\mathbf{R}_L(t)
\end{equation}
at time $t$, where
\begin{equation}
    \mathbf{H}
    =
    \begin{bmatrix}
    \alpha_{1,1}&\cdots&\alpha_{n+1}\\
    \vdots&\ddots&\vdots\\
    \alpha_{N,1}&\cdots&\alpha_{N+1}\\
    \end{bmatrix}
    \in \mathbb{R}^{N\times \left(n+1\right)}
    .
\end{equation}
\begin{theorem}\label{Lem1}
If initial positions of leaders satisfy rank condition \eqref{LeadersRank}, followers' in-neighbors are obtained by \eqref{opt33} and \eqref{opt34}, and followers' communication weights are assigned using relation \eqref{comweightss}, then, the following properties hold:
\begin{subequations}
\begin{equation}
\label{secondwl}
\mathbf{H}=-\mathbf{L}^{-1}\mathbf{L}_0=-
\begin{bmatrix}
-\mathbf{I}_{n+1}\\
\left(-\mathbf{I}_{n+1}+\mathbf{G}\right)^{-1}\mathbf{F}
\end{bmatrix}
,
\end{equation}
\begin{equation}
\label{y-ya}
t\in [t_0,t_f],\qquad    \mathbf{Y}_d(t)-\mathbf{Y}(t)=\left(\mathbf{I}_3\otimes\mathbf{L}\right)\left(\mathbf{Y}(t)-\mathbf{Y}_a(t)\right),
\end{equation}
\end{subequations}
where ``$\otimes $'' is the Kronecker product symbol, {\color{black}and where}
\begin{subequations}
\begin{equation}
\label{Y}
    \mathbf{Y}(t)=\mathrm{vec}\left(
    \begin{bmatrix}
    \mathbf{r}_1(t)&\cdots&\mathbf{r}_N(t)
    \end{bmatrix}
    ^T
    \right),
\end{equation}
\begin{equation}
\label{Yd}
    \mathbf{Y}_d(t)=\mathrm{vec}\left(
    \begin{bmatrix}
    \mathbf{r}_{1,d}(t)&\cdots&\mathbf{r}_{N,d}(t)
    \end{bmatrix}
    ^T
    \right),
\end{equation}
\end{subequations}
{\color{black}aggregate} actual, local desired, and global desired positions of all agents at time $t$, and 
\begin{equation}
\label{L0}
    \mathbf{L}_0=\begin{bmatrix}
\mathbf{I}_{n+1}\\
\mathbf{0}
\end{bmatrix}
\in \mathbb{R}^{3N\times 3(n+1)}.
\end{equation}
\end{theorem}

\subsection{MQS Collective Dynamics {\color{black}and Coordination Control}}
\label{MAS Collective Dynamics1}
{\color{black}{\color{black}A  feedback controller needs to be designed}, for each individual quacopter $i\in \mathcal{V}$, to stably track the reference trajectory $\mathbf{r}_{i,d}(t)$ at any time $t\in [t_0,t_f]$. {\color{black}
\begin{definition}\label{liederivative}
Let $y:\mathbb{R}^{p}\rightarrow \mathbb{R}$ and $\mathbf{f}:\mathbb{R}^{p}\rightarrow \mathbb{R}^p$ be smooth functions. The Lie derivative $y$ with respect to $\mathbf{f}$ is defined as follows:
\[
L_{\mathbf{f}}y=\bigtriangledown y \mathbf{f}.
\]
\end{definition}
Let $\mathbf{g}_i\left(\mathbf{x}_i\right)=\begin{bmatrix}
\mathbf{g}_{1,i}\left(\mathbf{x}_i\right)&\cdots&\mathbf{g}_{4,i}\left(\mathbf{x}_i\right)
\end{bmatrix}$ and $\mathbf{u}_i=\begin{bmatrix}
u_{1,i}&\cdots&u_{4,i}
\end{bmatrix}^T$ where $\mathbf{g}_{1,i}$ through $\mathbf{g}_{4,i}$ are the columns of matrix $\mathbf{g}_i$, and $u_{1,i}=p_i$, $u_{2,i}=\tau_{\phi,i}$, $u_{3,i}=\tau_{\theta,i}$, and $u_{4,i}=\tau_{\psi,i}$. By considering Definition \ref{liederivative} and defining $\mathbf{r}_i=\begin{bmatrix}
x_i~y_i~z_i
\end{bmatrix}^T$ as the output of quadcopter $i\in \mathcal{V}$, we can write
\begin{equation}
\label{drawbackeq}
    q\in \{x,y,z\},~i\in \mathcal{V},\qquad \ddot{q}_i=L_{\mathbf{f}_i}^2q_i+\sum_{h=1}^4L{\mathbf{g}_{_{h,i}}}L_{\mathbf{f}_i}q_iu_{h,i}.
\end{equation}
where $L{\mathbf{g}_{_{2,i}}}L_{\mathbf{f}_i}q_i=L{\mathbf{g}_{_{3,i}}}L_{\mathbf{f}_i}q_i=L{\mathbf{g}_{_{4,i}}}L_{\mathbf{f}_i}q_i=0$. Thus, $u_{2,i}$, $u_{3,i}$, and $u_{4,i}$ do not appear on the right-hand side of Eq. \eqref{drawbackeq}. 

To overcome this issue, we extend the quadcopter dynamics \eqref{generalnonlineardynamics} to 
\begin{equation}
\label{extendednonlineardynamicsquadcopter}
\begin{cases}
    \tilde{\mathbf{x}}_i=\tilde{\mathbf{f}}_i\left(\tilde{\mathbf{x}}_i\right)+\tilde{\mathbf{g}}_{i}\left(\tilde{\mathbf{x}}_i\right)\tilde{\mathbf{u}}_i\\
    \mathbf{r}_i=\tilde{\mathbf{C}}\tilde{\mathbf{x}}_i
\end{cases}
,
\end{equation}
where $\tilde{\mathbf{x}}_i=\begin{bmatrix}{\mathbf{x}}_i^T&p_i&\dot{p}_i\end{bmatrix}^T$, $\tilde{\mathbf{u}}_i=\begin{bmatrix}u_{p,i}&\tau_{\phi,i}&\tau_{\theta,i}&\tau_{\psi,i}\end{bmatrix}^T$, {\color{black}$\tilde{\mathbf{C}}_i=\begin{bmatrix}
    \mathbf{I}_3&\mathbf{0}_{3\times 11}
    \end{bmatrix}$,}
   \[
    {\color{black}
   \resizebox{0.99\hsize}{!}{%
$
\tilde{\mathbf{f}}_i\left(\tilde{\mathbf{x}}_i\right)=
\begin{bmatrix}
{\mathbf{f}}_i\left({\mathbf{x}}_i\right)\\
\dot{p}_i\\
0\\
\end{bmatrix}+\begin{bmatrix}
\mathbf{0}_{3\times 1}\\
{p_i\over m_i}\hat{\mathbf{k}}_{b,i}\\
\mathbf{0}_{8\times 1}\\
\end{bmatrix},
~\mathrm{and}~\tilde{\mathbf{g}}_i\left(\tilde{\mathbf{x}}_i\right)=
\begin{bmatrix}
\mathbf{0}_{9\times 1}&\mathbf{0}_{9\times 3}\\
\mathbf{0}_{3\times 1}&\mathbf{J}_i^{-1}\\
0&\mathbf{0}_{1\times 3}\\
1&\mathbf{0}_{1\times 3}\\
\end{bmatrix}
\begin{bmatrix}
u_{p,i}\\
\tau_{\phi,i}\\
\tau_{\theta,i}\\
\tau_{\psi,i}\\
\end{bmatrix}
.
$
}}
\]
Define $\tilde{\mathbf{g}}_i\left(\mathbf{x}_i\right)=\begin{bmatrix}
\tilde{\mathbf{g}}_{1,i}\left(\tilde{\mathbf{x}}_i\right)&\cdots&\tilde{\mathbf{g}}_{4,i}\left(\tilde{\mathbf{x}}_i\right)
\end{bmatrix}$ and $\tilde{\mathbf{u}}_i=\begin{bmatrix}
\tilde{u}_{1,i}&\cdots&\tilde{u}_{4,i}
\end{bmatrix}^T$ where $\tilde{\mathbf{g}}_{1,i}$ through $\tilde{\mathbf{g}}_{4,i}$ are the columns of matrix $\tilde{\mathbf{g}}_i$, and $\tilde{u}_{1,i}=u_{p,i}$, $\tilde{u}_{2,i}=\tau_{\phi,i}$, $\tilde{u}_{3,i}=\tau_{\theta,i}$, and $\tilde{u}_{4,i}=\tau_{\psi,i}$. Here, we can write
\begin{equation}
\label{59mainextended}
    q\in \{x,y,z\},~i\in \mathcal{V},\qquad \ddddot{q}_i=L_{\tilde{\mathbf{f}}_i}^{4}q_i+\sum_{h=1}^4L_{\tilde{\mathbf{g}}_{_{h,i}}}L_{\tilde{\mathbf{f}}_i}^{3}q_i\tilde{u}_{h,i},
\end{equation}
where $L_{\tilde{\mathbf{g}}_{_{h,i}}}L_{\tilde{\mathbf{f}}_i}^{3}q_i\neq0$ for $h=1,\cdots, 4$ and $q\in \{x,y,z\}$. Therefore, the extended dynamics \eqref{extendednonlineardynamicsquadcopter} is input-output linearizable. By defining the state transformation $\tilde{\mathbf{x}}_i\rightarrow \left(\mathbf{z}_i,\bf{\zeta}_i\right)$, the extended dynamics \eqref{extendednonlineardynamicsquadcopter} is converted to the following {\color{black}internal} and {\color{black}external} dynamics:
\begin{subequations}
\begin{equation}
\dot{\bf{\psi}}_i=\begin{bmatrix}
0&1\\
0&0
\end{bmatrix}
{\bf{\psi}}_i+\begin{bmatrix}
0\\
1
\end{bmatrix}
u_{\psi,i},
\end{equation}
\begin{equation}
\label{linearizeddynamicsquadcopteri}
\begin{split}
    \dot{\mathbf{z}}_i=\mathbf{A}_i\mathbf{z}_i+\mathbf{B}_i\mathbf{s}_i,
\end{split}
\end{equation}
\end{subequations}
where $\mathbf{z}_i=\begin{bmatrix}{\mathbf{r}}_i^T&\dot{\mathbf{r}}_i^T&\ddot{\mathbf{r}}_i{\color{black}^T}&\dddot{\mathbf{r}}_i^T\end{bmatrix}^T$ and $\bf{\zeta}_i=\begin{bmatrix}
\psi_i&\dot{\psi}_i
\end{bmatrix}^T$ are the state vectors of the internal and external dynamics, respectively, $\mathbf{A}_i=\begin{bmatrix}
    \mathbf{0}_{9\times 3}&\mathbf{I}_9\\
    \mathbf{0}_{3\times 9}&\mathbf{0}_{3\times 9}
    \end{bmatrix}$, and $\mathbf{B}_i=\begin{bmatrix}
    \mathbf{0}_{9\times 3}\\
    \mathbf{I}_{3}
    \end{bmatrix}$. 
    
    \subsubsection{Feedback-Linearization Control Design }
    Define $\mathbf{v}_i=\begin{bmatrix}
    \mathbf{s}_i^T&u_{\psi,i}
    \end{bmatrix}^T$ as the vector aggregating the control inputs of the external and internal dynamics, respectively. If quadcopter $i\in \mathcal{V}$ is modeled by dynamics \eqref{extendednonlineardynamicsquadcopter}, then $\mathbf{v}_i$ and $\mathbf{u}_i$ are related by
    \begin{equation}
        \mathbf{v}_i=\mathbf{M}_i\mathbf{u}_i+\mathbf{N}_i,
    \end{equation}
    where
    \begin{subequations}
     \begin{equation}
    \resizebox{0.99\hsize}{!}{%
$
        \mathbf{M}_i=\begin{bmatrix}
        L_{\tilde{\mathbf{g}}_{_{1,i}}}L_{\tilde{\mathbf{f}}_i}^{3}x_i& L_{\tilde{\mathbf{g}}_{_{2,i}}}L_{\tilde{\mathbf{f}}_i}^{3}x_i& L_{\tilde{\mathbf{g}}_{_{3,i}}}L_{\tilde{\mathbf{f}}_i}^{3}x_i& L_{\tilde{\mathbf{g}}_{_{4,i}}}L_{\tilde{\mathbf{f}}_i}^{3}x_i\\
        L_{\tilde{\mathbf{g}}_{_{1,i}}}L_{\tilde{\mathbf{f}}_i}^{3}y_i& L_{\tilde{\mathbf{g}}_{_{2,i}}}L_{\tilde{\mathbf{f}}_i}^{3}y_i& L_{\tilde{\mathbf{g}}_{_{3,i}}}L_{\tilde{\mathbf{f}}_i}^{3}y_i& L_{\tilde{\mathbf{g}}_{_{4,i}}}L_{\tilde{\mathbf{f}}_i}^{3}y_i\\
        L_{\tilde{\mathbf{g}}_{_{1,i}}}L_{\tilde{\mathbf{f}}_i}^{3}z_i& L_{\tilde{\mathbf{g}}_{_{2,i}}}L_{\tilde{\mathbf{f}}_i}^{3}z_i& L_{\tilde{\mathbf{g}}_{_{3,i}}}L_{\tilde{\mathbf{f}}_i}^{3}z_i& L_{\tilde{\mathbf{g}}_{_{4,i}}}L_{\tilde{\mathbf{f}}_i}^{3}z_i\\
        1&0&0&0
        \end{bmatrix}
        =
        \left[
        \begin{array}{cc}
            \multicolumn{2}{c}{{1\over m_i} \mathbf{O}_{1,i}\mathbf{O}_{3,i}}\\
            1 &\mathbf{0}_{1\times 3} 
        \end{array}
        \right],
        $
        }
    \end{equation}
    \begin{equation}
        \mathbf{N}_i=\begin{bmatrix}
        L_{\tilde{\mathbf{f}}_i}^{4}x_i&
       L_{\tilde{\mathbf{f}}_i}^{4}y_i&
        L_{\tilde{\mathbf{f}}_i}^{4}z_i&
        0
        \end{bmatrix}
        ^T
        =
        \left[
        \begin{array}{cc}
          {1\over m_i}  \left(\mathbf{O}_{1,i}\mathbf{O}_{4,i}+\mathbf{O}_{2,i}\right)\\
            0
        \end{array}
        \right]
    \end{equation}
    \end{subequations}
    and $\mathbf{O}_{1,i}$ through $\mathbf{O}_{4,i}$ are defined in Appendix \ref{Time Derivatives of the Quadcopter Thrust Force}.
    
    We choose
\[
u_{\psi_i}=-k_{1,\psi,i}\dot{\psi}_i-k_{2,\psi,i}\psi_i,
\]
 $k_{1,\psi,i}>0$ and $k_{2,\psi,i}>0$. Therefore, $\psi_i(t)$ asymptotically converges to $0$. We also choose
\begin{equation}
\forall i\in \mathcal{V} \qquad    \mathbf{s}_i=-k_{1,i}\dddot{\mathbf{\mathbf{r}}}_i-k_{2,i}\ddot{\mathbf{\mathbf{r}}}_i-k_{3,i}\dot{\mathbf{\mathbf{r}}}_i+k_{4,i}\left(\mathbf{r}_{d,i}-{\mathbf{\mathbf{r}}}_i\right),
\end{equation}
where $k_{1,i}$ through $k_{4,i}$ are selected for every quadcopter $i\in \mathcal{V}$ such the stability of the MQS collective dynamics is ensured. A condition for stability of the MQS collective coordination is provided in Section \ref{MAS Collective Dynamics2}.
}

}

\subsubsection{{\color{black}MQS External Dynamics and Stability Analysis}}
\label{MAS Collective Dynamics2}

The external dynamics of the {\color{black}MQS} is given by
\begin{equation}
\begin{cases}
\dot{\mathbf{Z}}=\mathbf{A}_{\mathrm{SYS}}\mathbf{Z}+\mathbf{B}_{\mathrm{SYS}}\mathbf{S}\\
\mathbf{Y}=\mathbf{C}_{\mathrm{SYS}}\mathbf{Z}
\end{cases},
\end{equation}
where $\mathbf{Y}=\mathrm{vec}\left(\begin{bmatrix}
\mathbf{r}_1&\cdots&\mathbf{r}_N
\end{bmatrix}^T\right)$, $\mathbf{Z}=\begin{bmatrix}
\mathbf{z}_1^T&\cdots&{\color{black}\mathbf{z}}_N^T
\end{bmatrix}^T$, $\mathbf{S}=\begin{bmatrix}
\mathbf{s}_1^T&\cdots&\mathbf{s}_N^T
\end{bmatrix}^T$, $\mathbf{C}_{\mathrm{SYS}}\in \mathbb{R}^{3N\times 12N}$, $\mathbf{A}_{\mathrm{SYS}}=\mathrm{diag}\left(\mathbf{A}_1,\cdots,\mathbf{A}_N\right)\in \mathbb{R}^{12N\times 12N}$, and $\mathbf{B}_{\mathrm{SYS}}=\mathrm{diag}\left(\mathbf{B}_1,\cdots,\mathbf{B}_N\right)\in \mathbb{R}^{12N\times 3N}$. 
 Given the local desired trajectory definition in \eqref{LocalDesiredEquation}, 
\begin{equation}
\begin{bmatrix}
 \mathbf{Y}_d(t)\\
  \dot{\mathbf{Y}}_d(t)\\
   \ddot{\mathbf{Y}}_d(t)\\
    \dddot{\mathbf{Y}}_d(t)
\end{bmatrix}
   =\left(\mathbf{I}_{12}\otimes \mathbf{W}\right)\begin{bmatrix}
 \mathbf{Y}(t)\\
  \dot{\mathbf{Y}}(t)\\
   \ddot{\mathbf{Y}}(t)\\
    \dddot{\mathbf{Y}}(t)
\end{bmatrix}+\left(\mathbf{I}_{12}\otimes\mathbf{L}_0\right)
\begin{bmatrix}
 \mathbf{R}_L(t)\\
  \dot{\mathbf{R}}_L(t)\\
   \ddot{\mathbf{R}}_L(t)\\
    \dddot{\mathbf{R}}_L(t)
\end{bmatrix}
\end{equation}
where  $\mathbf{I}_{12}\in \mathbb{R}^{12\times 12}$ is the identity matrix; $\mathbf{W}${\color{black},} $\mathbf{R}_L$, and $\mathbf{L}_0$ were previously defined in \eqref{WW}, \eqref{RL}, and \eqref{L0}, respectively. 
the external dynamics of the MQS can be expressed as follows:
\begin{equation}
\label{mainlinearized}
\dfrac{d}{dt}\left(
\begin{bmatrix}
\mathbf{Y}\\
\dot{\mathbf{Y}}\\
\ddot{\mathbf{Y}}\\
\dddot{\mathbf{Y}}\\
\end{bmatrix}
\right)
=\mathbf{A}_{\mathrm{MQS}}
\begin{bmatrix}
\mathbf{Y}\\
\dot{\mathbf{Y}}\\
\ddot{\mathbf{Y}}\\
\dddot{\mathbf{Y}}\\
\end{bmatrix}
+
\mathbf{B}_{\mathrm{MQS}}
\begin{bmatrix}
\mathbf{R}_L\\
\dot{\mathbf{R}}_L\\
\ddot{\mathbf{R}}_L\\
\dddot{\mathbf{R}}_L\\
\end{bmatrix}
,
\end{equation}
where
\begin{subequations}
\begin{equation}
\mathbf{A}_{\mathrm{MQS}}=
    \begin{bmatrix}
\mathbf{0}&\mathbf{I}_{3N}&\mathbf{0}&\mathbf{0}\\
\mathbf{0}&\mathbf{0}&\mathbf{I}_{3N}&\mathbf{0}\\
\mathbf{0}&\mathbf{0}&\mathbf{0}&\mathbf{I}_{3N}\\
\mathbf{I}_3\otimes\left( \mathbf{K}_4 \mathbf{L}\right)&\mathbf{I}_3\otimes \left(\mathbf{K}_3 \mathbf{L}\right)&\mathbf{I}_3\otimes\left( \mathbf{K}_2\mathbf{L}\right)&\mathbf{I}_3\otimes \left( \mathbf{K}_1 \mathbf{L}\right)
\end{bmatrix}
,
\end{equation}
\begin{equation}
\resizebox{0.99\hsize}{!}{%
$
\mathbf{B}_{\mathrm{MQS}}=
    \begin{bmatrix}
\mathbf{0}&\mathbf{0}&\mathbf{0}&\mathbf{0}\\
\mathbf{0}&\mathbf{0}&\mathbf{0}&\mathbf{0}\\
\mathbf{0}&\mathbf{0}&\mathbf{0}&\mathbf{0}\\
\mathbf{I}_3\otimes\left( \mathbf{K}_{4} \mathbf{L}_0\right)&\mathbf{I}_3\otimes \left(\mathbf{K}_{3} \mathbf{L}_0\right)&\mathbf{I}_3\otimes\left( \mathbf{K}_2\mathbf{L}_0\right)&\mathbf{I}_3\otimes \left( \mathbf{K}_1 \mathbf{L}_0\right)
\end{bmatrix}
,
$
}
\end{equation}
\begin{equation}
    j=1,2,3,4,\qquad \mathbf{K}_j=\mathrm{diag}\left(k_{j,1},\cdots,k_{j,N}\right),
\end{equation}
\end{subequations}
and $\mathbf{I}_{3N}\in \mathbb{R}^{3N\times 3N}$ is the identity matrix.
Note that control gains $k_{j,i}$ ($i\in \mathcal{V}$ and $j=1,2,3,4$) are selected such that roots of the characteristic equation
\begin{equation}
\label{CharEq}
    \left|s^4\mathbf{I}+s^3\mathbf{K}_1+s^2\mathbf{K}_2+s\mathbf{K}_3+\mathbf{K}_4\right|=0
\end{equation}
are all located {\color{black}in the open left half of complex plane.} The block diagram of the MQS control system is shown in Fig. \ref{Detecton2D}. 
\vspace{-.5cm}

\subsection{Inter-Agent Collision Avoidance and {\color{black}Quadcopter} Containment}
\label{Inter-Agent Cion Avoidance}
To ensure inter-agent collision avoidance and quadcopter containement, safety conditions \eqref{eqreferror12}, \eqref{intagcol}, and \eqref{quadcontainmentttttt} must be satisfied. Conditions \eqref{intagcol} and \eqref{quadcontainmentttttt} can be guaranteed by defining admissible affine transformation features as discussed in Section \ref{Problem 1: Affine Transformation Definition}. {\color{black}To ensure \eqref{eqreferror12}, we assume that $k_{j,1}$, $k_{j,2}$, $\cdots$, and $k_{j,2}$ ($\forall j\in \{1,\cdots,4\}$) are selected such that the roots of the Characteristic Eq. \eqref{CharEq} are all placed {\color{black}in the open left half of complex plane}. Then,  we guarantee the  safety condition \eqref{eqreferror12} by choosing a sufficiently-large {\color{black}maneuver duration, i.e.,} $T$.} 




{\color{black}Define $\mathbf{E}=\mathbf{Y}(t)-\mathbf{Y}_a(t)$ {\color{black}as} the error vector.} Per Theorem \ref{Lem1}, $\mathbf{Y}_d(t)-\mathbf{Y}(t)=\left(\mathbf{I}_3\otimes\mathbf{L}\right)\mathbf{E}$; thus, the error dynamics becomes
\begin{equation}
\label{errorlinearized}
\resizebox{0.99\hsize}{!}{%
$
\dfrac{d}{dt}\left(
\begin{bmatrix}
\mathbf{E}^T&
\dot{\mathbf{E}}^T&
\ddot{\mathbf{E}}^T&
\dddot{\mathbf{E}}^T
\end{bmatrix}
^T
\right)
=\mathbf{A}_{\mathrm{MQS}}
\begin{bmatrix}
\mathbf{E}^T&
\dot{\mathbf{E}}^T&
\ddot{\mathbf{E}}^T&
\dddot{\mathbf{E}}^T
\end{bmatrix}
^T
+
\mathbf{V}_{\mathrm{MQS}}
,
$
}
\end{equation}
where 
\begin{equation}
\mathbf{V}_{\mathrm{MQS}}(t)=
\begin{bmatrix}
\mathbf{0}&
\mathbf{0}&
\mathbf{0}&
\mathbf{I}_3\otimes \mathbf{H}^T\\
\end{bmatrix}
^T
{\color{black}\ddddot{\mathbf{R}}}_L(t).
\end{equation}
Therefore,
\begin{equation}
\label{BB}
{\color{black}
\begin{bmatrix}
\mathbf{E}(t)\\
\dot{\mathbf{E}}(t)\\
\ddot{\mathbf{E}}(t)\\
\dddot{\mathbf{E}}(t)\\
\end{bmatrix}
}=
    \mathrm{e}^{\mathbf{A}_{\mathrm{MQS}}{\left(t-t_0\right)}}
    {\color{black}
\begin{bmatrix}
\mathbf{E}(t_0)\\
\dot{\mathbf{E}}(t_0)\\
\ddot{\mathbf{E}}(t_0)\\
\dddot{\mathbf{E}}(t_0)\\
\end{bmatrix}
}
    +\int_{t_0}^t\mathrm{e}^{\mathbf{A}_{\mathrm{SYS}}\left(t-\tau\right)}{\mathbf{V}_{\mathrm{MQS}}\left(\tau\right)}\mathrm{d}\tau{\color{black}.}
\end{equation}
Note that 
\begin{equation}
t\in[t_0,t_f],\qquad   \|\mathbf{r}_i(t)-\mathbf{r}_{i,a}(t)\|_2^2=\mathbf{E}^T(t)\mathbf{C}_i^T\mathbf{C}_i\mathbf{E}(t),
\end{equation}
where $\mathbf{C}_{i}=[{\mathbf{C}_i}_{_{lh}}]\in {\mathbb{R}^{3\times 12 N}}$ {\color{black}is a matrix with
the $(l,h)$ element of which
is given by}
\[
\resizebox{0.99\hsize}{!}{%
$
{\mathbf{C}_i}_{_{lh}}=
\begin{cases}
    1&\left(l=1\wedge h=i\right)\vee \left(l=2\wedge h=i+4N\right)\vee \left(l=3\wedge h=i+8N\right)\\
   0&\mathrm{otherwise}
\end{cases}
.$
}
\]
{\color{black}
 \begin{theorem}\label{theorem9999}
Assume the initial condition $\mathbf{E}(t_0)$, $\dot{\mathbf{E}}(t_0)$, $\ddot{\mathbf{E}}(t_0)$, and $\dddot{\mathbf{E}}(t_0)$ are given such that the {\color{black}the trajectory of} \eqref{errorlinearized} satisfies
\[
\forall t\in [t_0,t_f],~\forall i\in \mathcal{V},\qquad \|\mathbf{r}_i(t)-\mathbf{r}_{i,a}(t)\|<\varrho \delta,
\]
where $0<\varrho<1$. {\color{black}Then,} there exists a $\tilde{t}_f>0$ such that \eqref{eqreferror12} {\color{black}holds}, if $t_f\geq \tilde{t}_f$. 

\end{theorem}

The inter-agent collision avoidance can be ensured by choosing $T\geq T^*$ where $T_k^*=t_f^*-t_0$ is assigned as the solution of the following constrained optimization problem:
\begin{equation}
\label{tffffffff}
    t_f^*=\argmin\left(t_f-t_0\right)
\end{equation}
subject to 
\begin{equation}
\label{safety}
t\in [t_0,t_f],\qquad \bigwedge_{i\in \mathcal{V}}\mathbf{E}^T(t)\mathbf{C}_i^T\mathbf{C}_i\mathbf{E}(t)\leq \delta^2,
\end{equation}
where $t_0$ is known. 

\begin{remark}
By satisfaction of constraint \eqref{eqreferror12}, deviation of every quadcopter from its global desired trajectory remains bounded at any time $t$, and thus, safety of the MQS affine transformation can be ascertained only by constraining eigenvalues of the deformation matrix $\mathbf{U}_D$, by conditions \eqref{intagcol} and \eqref{quadcontainmentttttt}. We can guarantee the satisfaction of the safety requirements \eqref{eqreferror12}, \eqref{intagcol}, and \eqref{quadcontainmentttttt} without constraining the total number of {\color{black}quadcopters} participating in an affine transformation coordination. Thus, our proposed multi-agent coordination approach {\color{black}is scalable to large values of $N$.} 
\end{remark}
}

\begin{figure}
 \centering
 \subfigure[]{\includegraphics[width=0.46\linewidth]{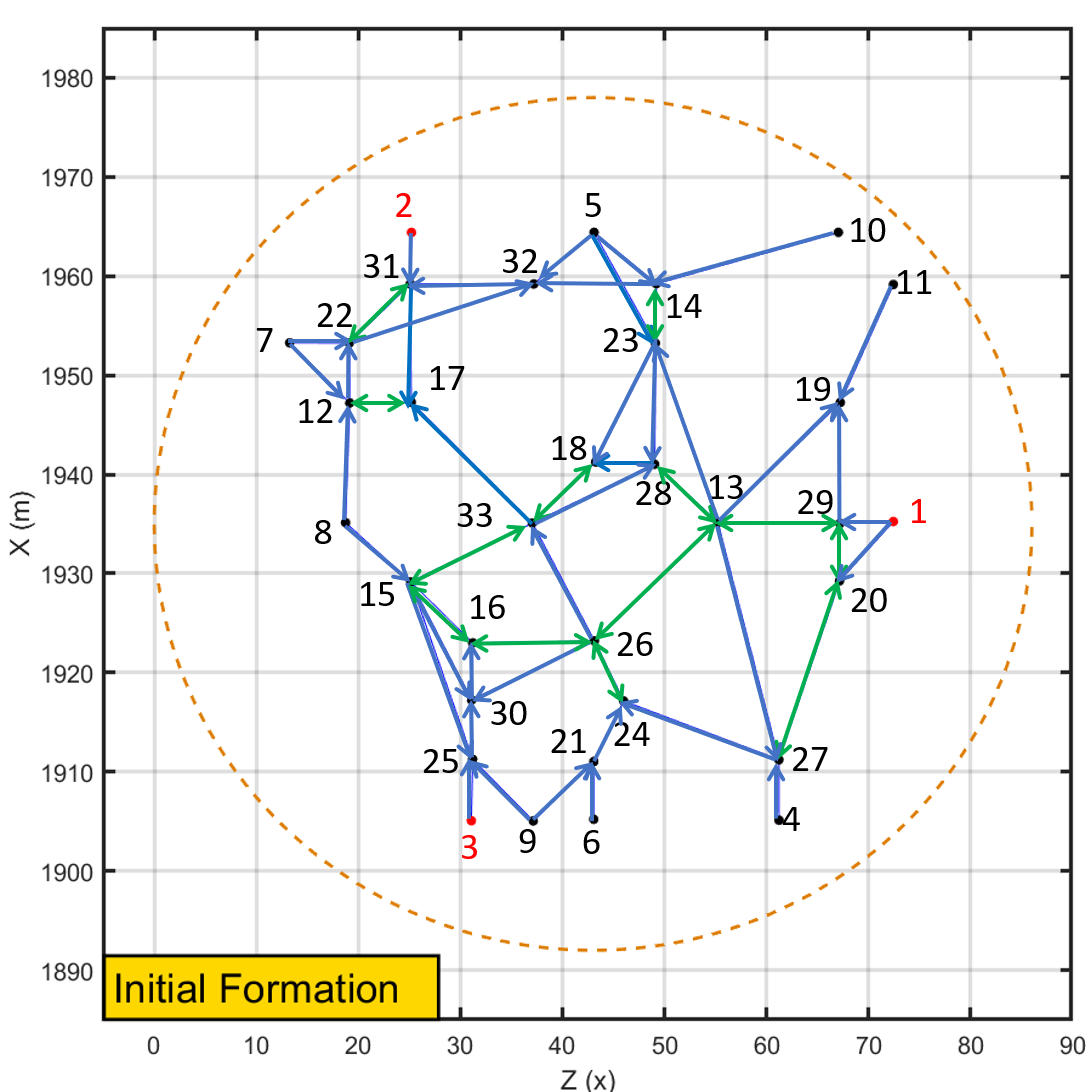}}
  \subfigure[]{\includegraphics[width=0.48\linewidth]{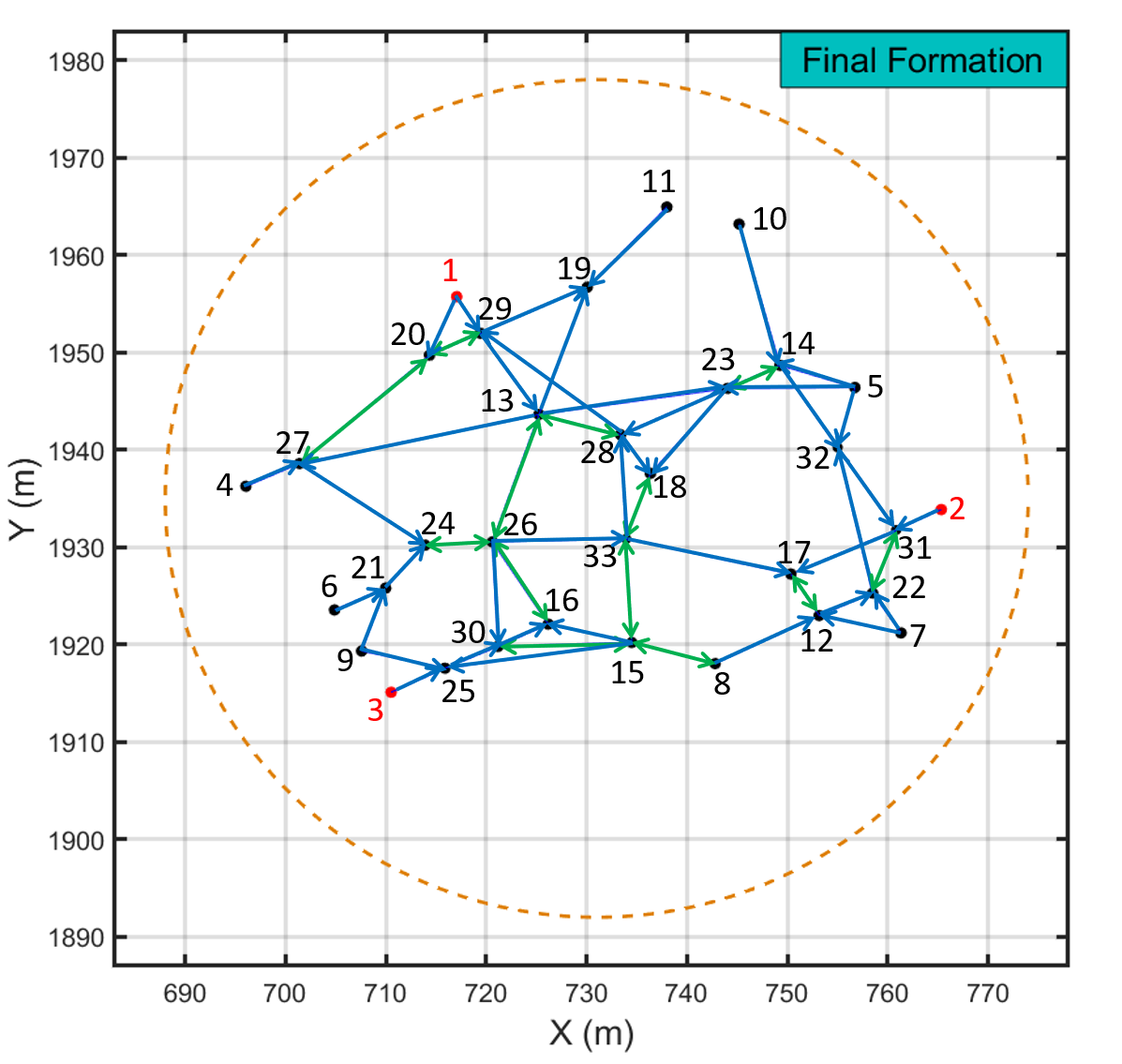}}
     \caption{(a,b) MQS initial and final formations.}
\label{Formation}
\end{figure}
\vspace{-.5cm}
\section{Simulation Results}\label{Simulation Results}
Consider an MQS consisting of $33$  quadcopters {\color{black}with the initial formation {\color{black}distributed in the $z-x$ plane} 
}  as shown in Fig. \ref{Formation} (a). {\color{black}For the initial configuration,}  $\lambda_{1,0}=\lambda_{2,0}=\lambda_{3,0}=1$, $\beta_{1,0}=\beta_{2,0}=0$, $\beta_{3,0}=\beta_{4,0}={\color{black}0~r}ad$, {\color{black}and $\bar{\mathbf{d}}_0=\begin{bmatrix}
1935&215&43
\end{bmatrix}^T$}. It is desired that the MQS ultimately {\color{black}achieves} the final configuration distributed in the $x-y$ plane as shown in Fig. \ref{Formation} (b) {by moving in an obstacle-laden environment shown in Fig. \ref{eigenvalues} (a)}. The final configuration is an affine transformation of the initial formation and characterized by the following features: $\lambda_{1,f}=\lambda_{3,f}=1$, $\lambda_{2,f}=-0.8$ $\beta_{1,f}=\beta_{2,f}{\color{black}=\beta_{3,f}}=\beta_{4,f}={\color{black}0~r}ad$, and {\color{black}$\bar{\mathbf{d}}_f=\begin{bmatrix}
731&1935&43
\end{bmatrix}^T$.} The shear deformation angles $\beta_5(t)=\beta_{5,0}=\beta_{5,f}$ and $\beta_6(t)=\beta_{6,0}=\beta_{6,f}$ are constant at any time $t$, where $\beta_{5,0}={\color{black}0~r}ad$ and {\color{black}$\beta_{6,0}=2.0735~rad$ are obtained by solving Eq. \eqref{beta5beta6}}. Given quadcopters' initial positions, followers' in-neighbors and communication weights are computed using the approach presented in Section \ref{Problem 1: Affine Transformation Definition}.
Note that $\mathcal{V}=\{1,\cdots,33\}$ can be expressed as $\mathcal{V}=\mathcal{V}_B\bigcup\mathcal{V}_I$, where $\mathcal{V}_B=\{1,\cdots,{\color{black}11}\}$ and $\mathcal{V}_I=\{{\color{black}12},\cdots,33\}$. Also, {\color{black}the} set $\mathcal{V}_L=\{1,2,3\}$ and $\mathcal{V}_F=\{4,\cdots,33\}$ define index numbers of leaders and followers'{\color{black},} respectively.
\vspace{-.3cm}
{\color{black}
\subsection{Safety Conditions}
\textbf{{Assignment of $\delta$:}}
Because $\lambda_{1,0}=\lambda_{1,f}=1$, and $\lambda_1(t)$ is defined by Eq. \eqref{ThetaDefinition} at any time $t\in [t_0,t_f]$,  $\lambda_1(t)=1$ at every time $t\in \left[0,t_f\right]$ ($t_0=0s$). Therefore, $\lambda_{\mathrm{min}}=1$ is considered as the lower limit of eigenvalue $\lambda_{1}$: $\lambda_1(t)\geq 1$, $\forall t\in \left[t_0,t_f\right]$. Given quadcopters' initial positions{\color{black},} $d_{\mathrm{min}}=2\left(\delta+\epsilon\right)=0.4387~m$ is computed using \eqref{dminnnnnnn}. It is assumed that every quadcopter is enclosed by a ball of radius $\epsilon=0.10m$, therefore, {\color{black}$\delta=0.11{\color{black}5~m}$}.

\textbf{Assignment of $\lambda_{\mathrm{max}}$:} Given quadcopters' initial positions, $d_{\mathrm{max}}=38.0555
m$ is obtained by Eq. \eqref{dmaxxx}. Given $\epsilon=0.1m$ and $\delta=0.11{\color{black}5~m}$, $\lambda_{\mathrm{max}}=1.1243$ is obtained by Eq. \eqref{lmax}. Therefore, collision avoidance and quadcopter containment are guaranteed, if the following inequalities are satisfied at any time $t\in [t_0,t_f]$: $\|\mathbf{r}_i(t)-\mathbf{r}_{i,a}(t)\|\leq 0.1$ for every quadcopter $i\in \mathcal{V}$ and $\lambda_1(t),\lambda_2(t),\lambda_3(t)\leq 1.1234$.

\textbf{Assignment of Travel Time $T$:} We choose $t_0=0s$, thus, a $t_f=T\geq T^*$ needs to be selected to ensure safety, where $T^*=776$ is obtained by solving Eq. \eqref{tffffffff}{\color{black}-\eqref{safety}}. For simulation, we choose $T=780s$.
}

\vspace{-.3cm}
\subsection{Plots}
\begin{figure*}
 \centering
 \subfigure[MQS affine transformation]{\includegraphics[width=0.24\linewidth]{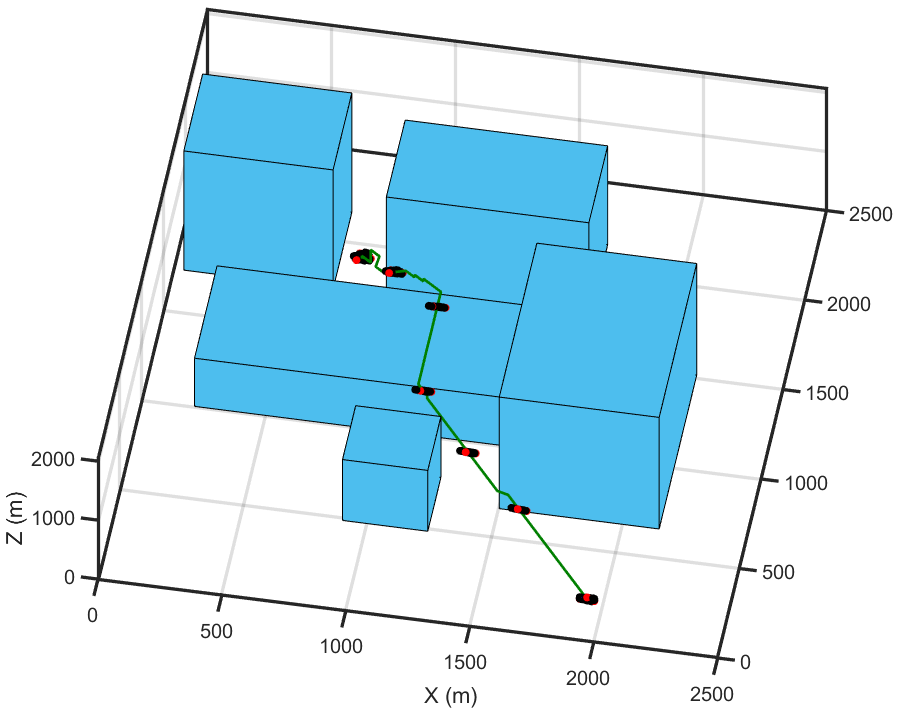}}
 \subfigure[$t=0,120,250s$]{\includegraphics[width=0.24\linewidth]{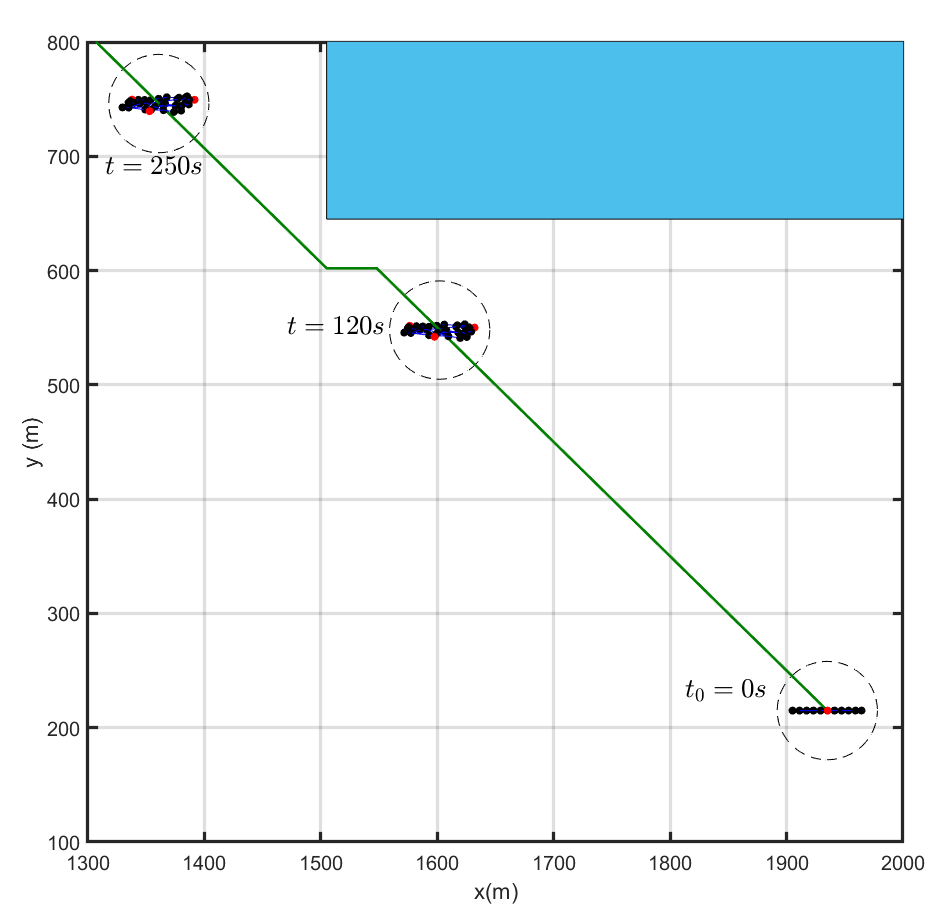}}
  \subfigure[$t=350,450s$]{\includegraphics[width=0.24\linewidth]{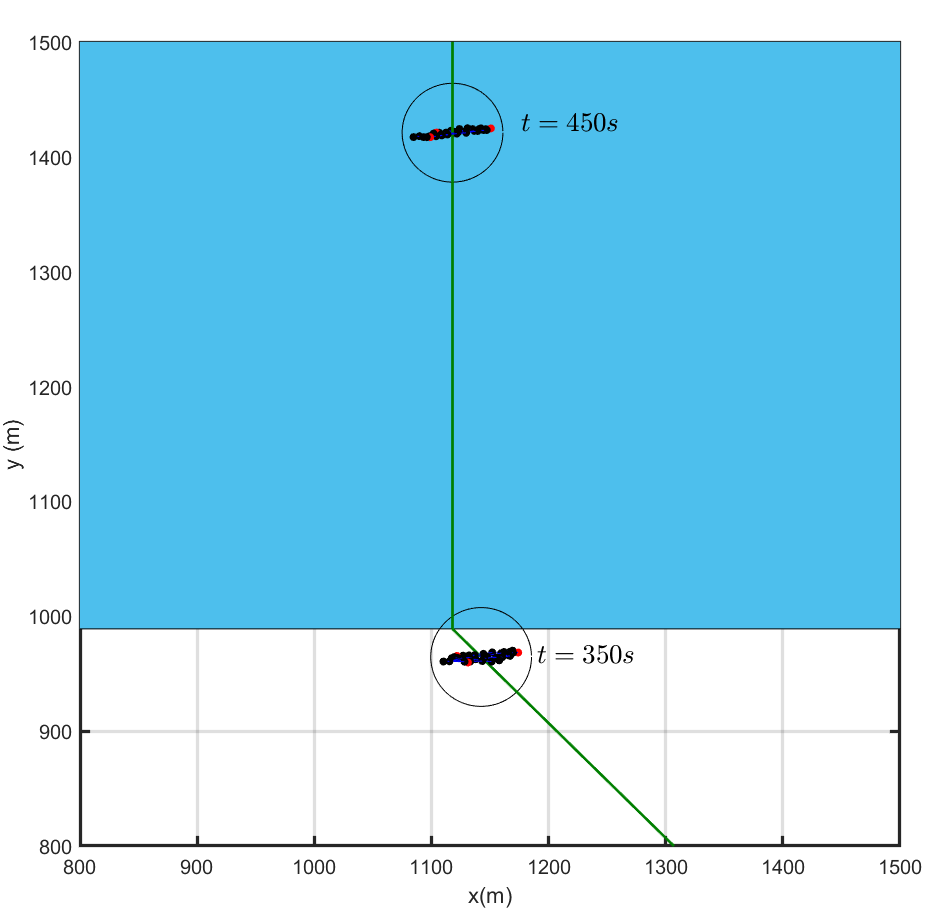}}
  \subfigure[$t=650,770s$]{\includegraphics[width=0.24\linewidth]{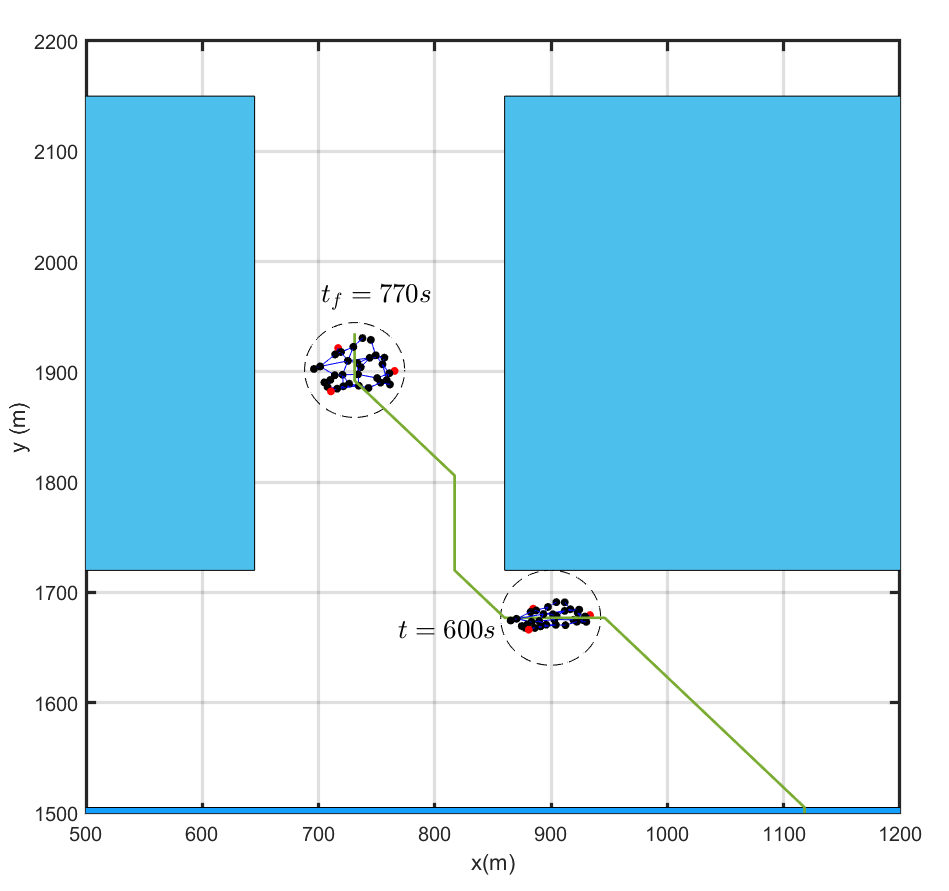}}
  \vspace{-.3cm}
     \caption{Affine transformation of the MQS in an obstacle-laden environment {\color{black}is illustrated in (a)}. Top view of MQS formations at sample times (b) $t=0,250,350s$, (c) $t=350,450s$, and (d) $t=650,770s$.}
\label{eigenvalues}
\end{figure*}

\begin{figure}
 \centering
 \subfigure[]{\includegraphics[width=0.95\linewidth]{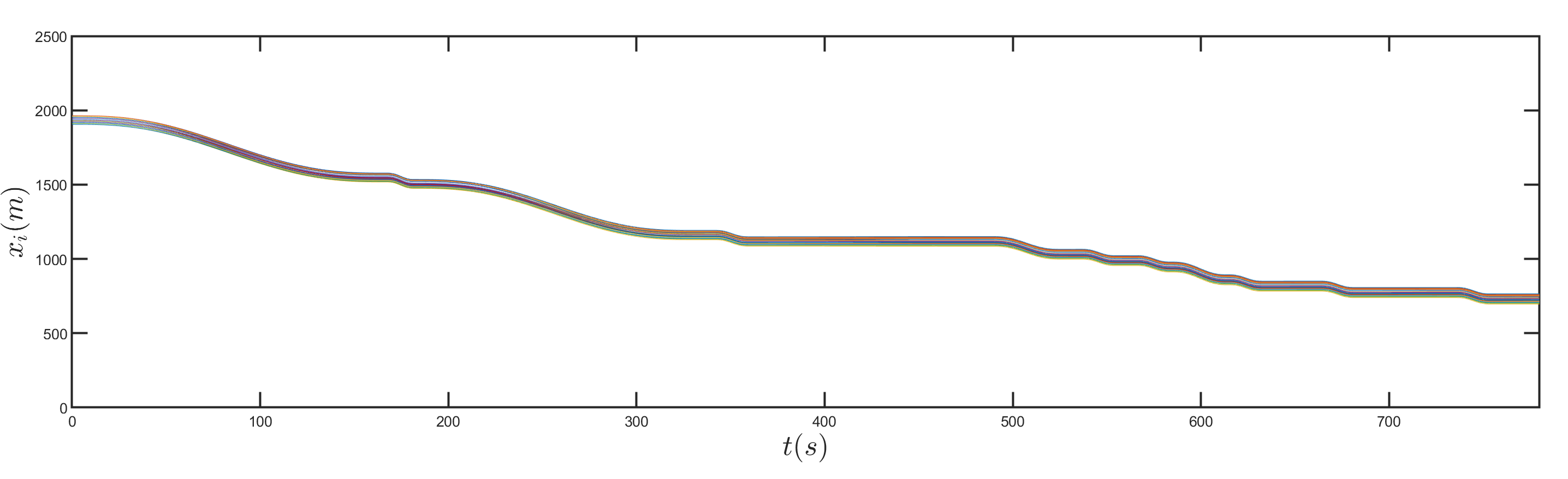}}
  \subfigure[]{\includegraphics[width=0.95\linewidth]{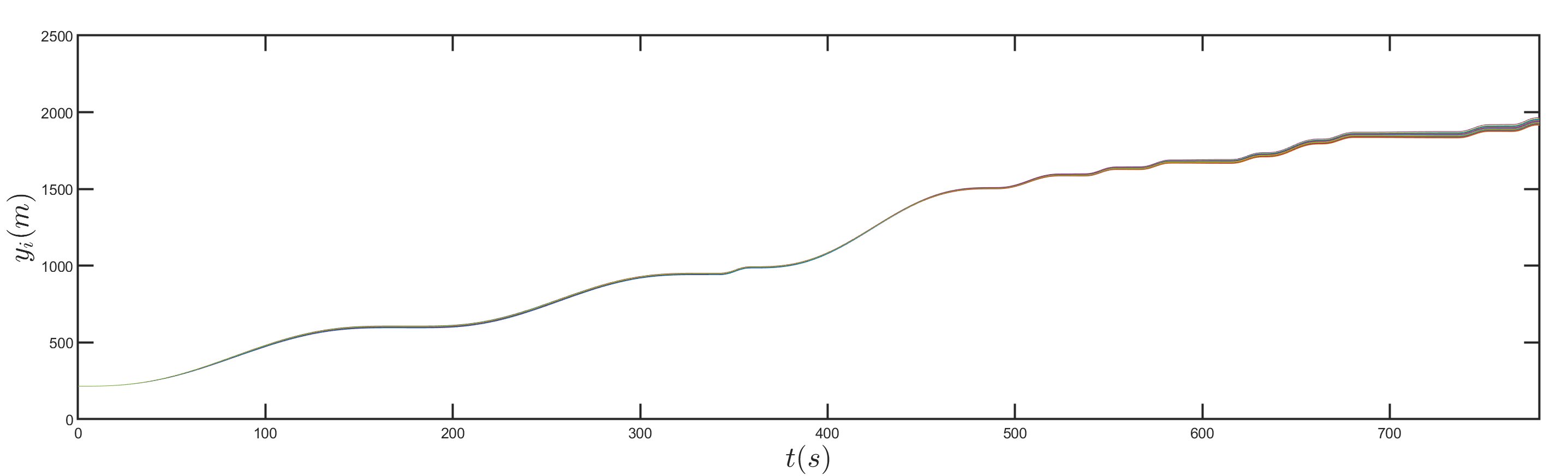}}
  \subfigure[]{\includegraphics[width=0.95\linewidth]{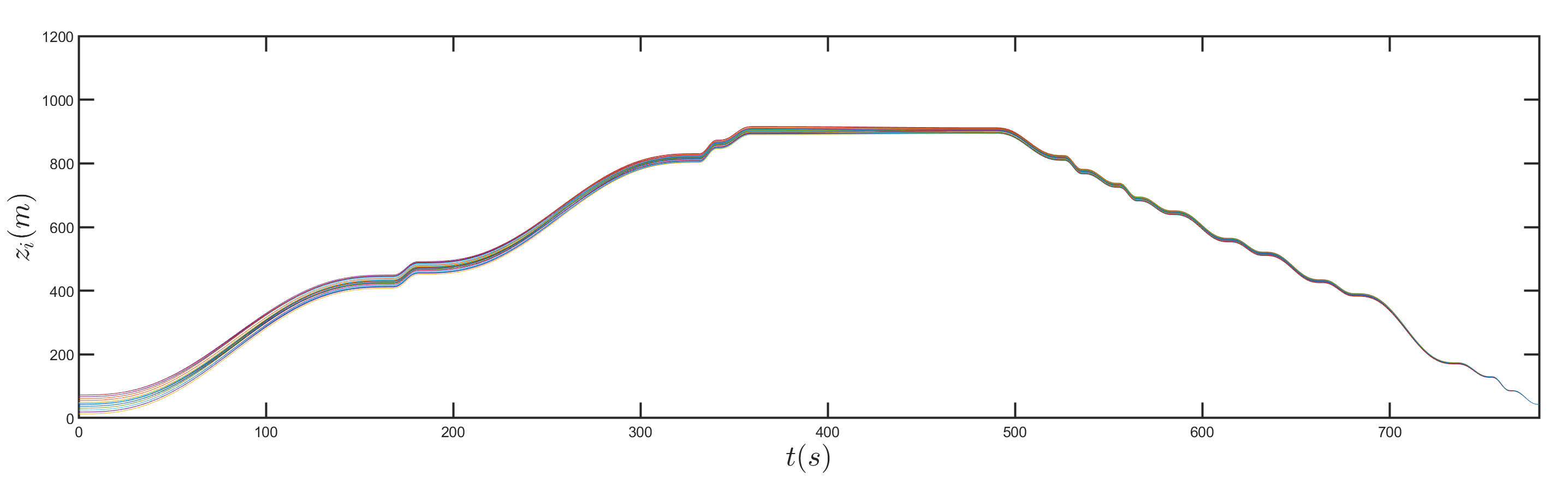}}
  \vspace{-.2cm}
     \caption{{\color{black}(a,b,c) $x$, $y$, and $z$} components of positions of all agents versus time $t$.}
\label{xyall}
\end{figure}

\begin{figure}[ht]
\centering
\includegraphics[width=3.3 in]{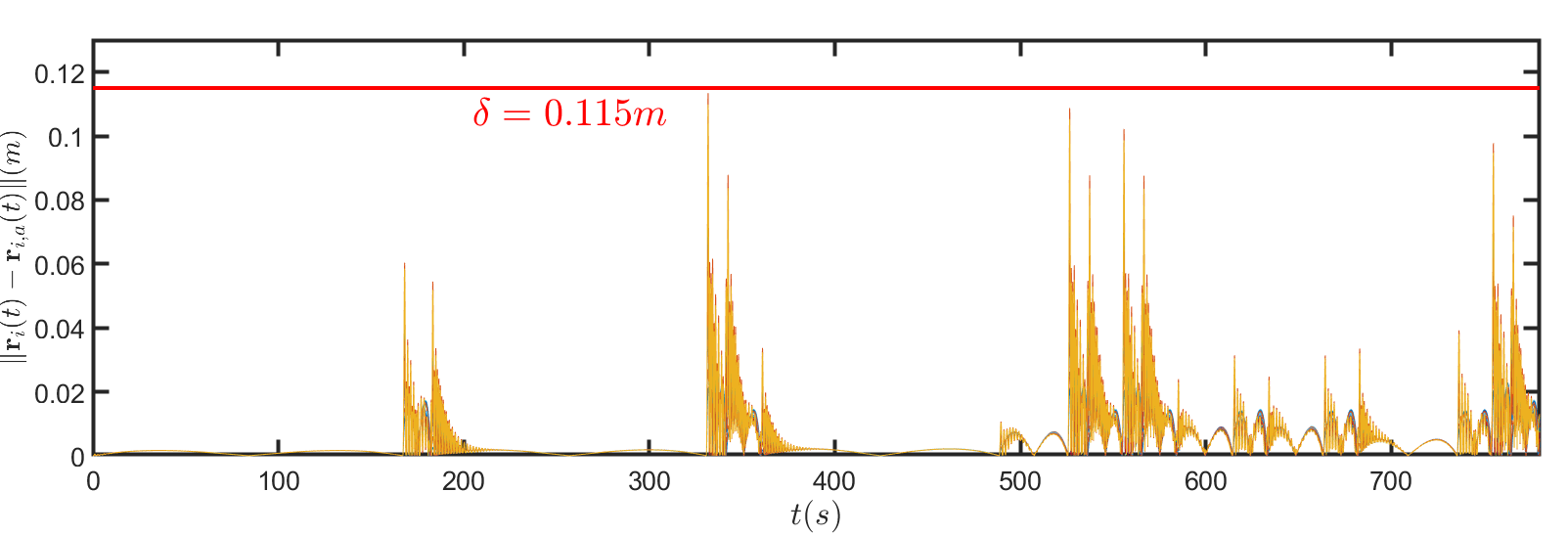}
\vspace{-.3cm}
\caption{Deviation of agents from global desired trajectories defined by an affine transformation.}
\label{error}
\end{figure}

\begin{figure}
 \centering
 \subfigure[]{\includegraphics[width=0.95\linewidth]{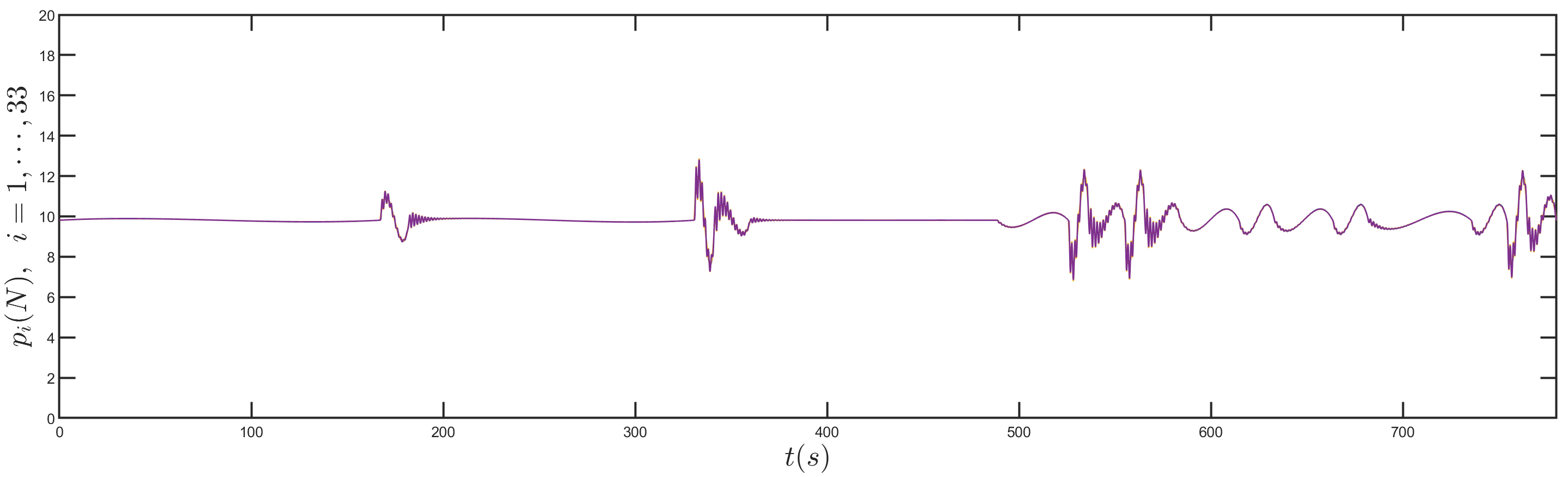}}
  \subfigure[]{\includegraphics[width=0.95\linewidth]{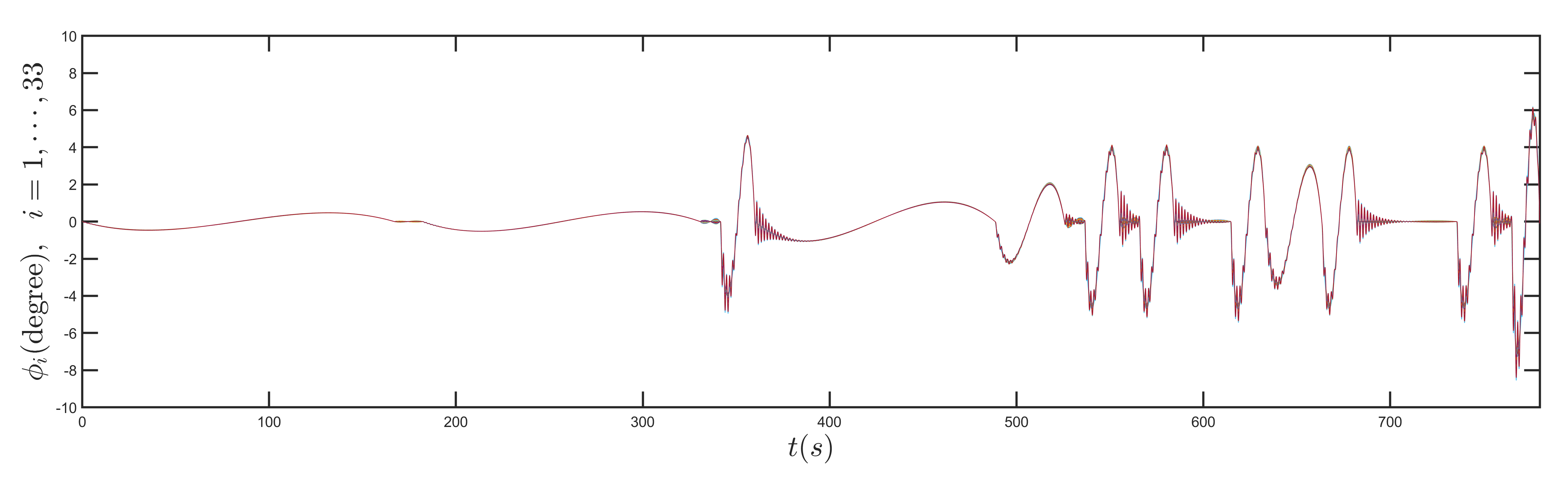}}
  \subfigure[]{\includegraphics[width=0.95\linewidth]{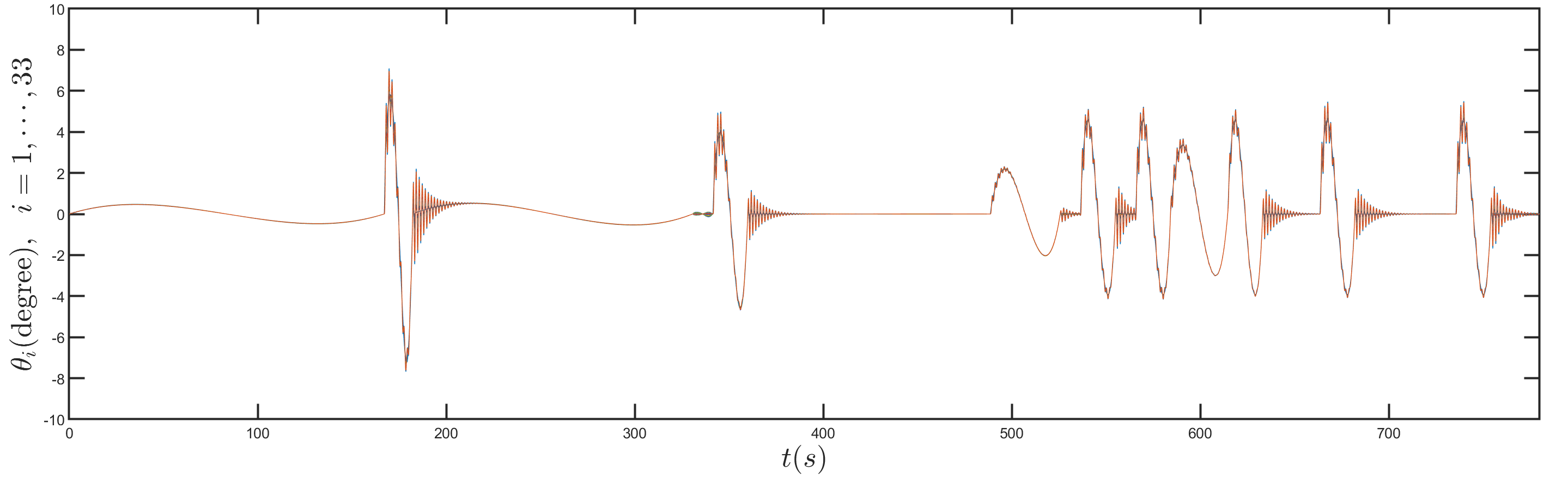}}
  \vspace{-.2cm}
     \caption{{\color{black}(a,b,c) Thrust force magnitude $p_i$, roll angle $\phi_i$, and pitch angle $\theta_i$ for every quadcopter $i$ versus time}.}
\label{forcesangles}
\end{figure}

{\color{black}In Fig. \ref{eigenvalues}(a), the optimal path of the containment ball $\mathcal{S}$ is shown by green. Furthermore, MQS formations are shown at sample times $0s$, $250s$, $350s$, $450s$, $650s$, and $770s$ in Figs. \ref{eigenvalues} (a-d). Note that Fig. \ref{eigenvalues} (b-d) plots the projections of the MQS formations on the $x-y$ plane at different sample times.  Additionally, $x$, $y$, and $z$ components of positions of all quadcopters are plotted versus time $t$ in Fig. \ref{xyall}. 
Fig. \ref{error} plots $\|\mathbf{r}_i-\mathbf{r}_{i,a}\|$ versus time for every agent $i\in \mathcal{V}$. It is seen that deviation of every quadcopter is less than  $\delta=0.11{\color{black}5~m}$ from its global desired position at any time $t\in[0,780]$. Figs. \ref{forcesangles} plot the thrust force magnitude $p_i$, roll angle $\phi_i$, and pitch angle $\theta_i$ for every quadcopter $i\in \mathcal{V}$ versus time.}

\vspace{-.3cm}
\section{Conclusion}\label{Conclusion}
This paper {\color{black}studied} the problem of large-scale affine transformation of an MQS in an obstacle-laden environment. By eigen-decomposition of the affine transformation, it {\color{black}was}  shown how a large-scale collective motion of an MQS can be safely planned such that inter-agent collision avoidance is avoided, quadcopter containment is guaranteed, and no quacopter hits an obstacle in an obstacle-laden environment. {\color{black}Similar to the previously proposed continuum deformation-based coordination approaches our method is scalable to coordination of a large numbers of vehicles, however, it allows to plan more efficient motions due to a more flexible form of the transformation being employed.}  A comprehensive comparison with other multi-agent coordination approaches  proposed in the literature and the development of further calibration/ tuning guidelines is beyond the scope of this paper and is left to future work.
Furthermore, the proposed affine transformation{\color{black}-based approach} paradigm improves the {\color{black}maneuverability} of the swarm coordination by relaxing thee restrictions {\color{black}in} the existing continuum deformation coordination {\color{black}approach}.

\vspace{-.3cm}
\bibliographystyle{IEEEtran}
\bibliography{reference}
\appendices
\section{Proofs}\label{Proofs}
\textbf{Proof of Proposition \ref{th1}:}
If Assumption \ref{Assume1} and rank condition \eqref{LeadersRank} are satisfied, {\color{black}} initial position of quadcopter $i\in \mathcal{V}$ can be expressed by the following linear combination{\color{black},}
\begin{equation}
\label{pprrooff}
    i\in \mathcal{V},\qquad \mathbf{r}_{i,0}-\mathbf{r}_{1,0}=\sum_{j=2}^{n+1}\alpha_{i,j}\left(\mathbf{r}_{j,0}-\mathbf{r}_{1,0}\right),
\end{equation}
where $\alpha_{i,2}$, $\cdots$, $\alpha_{i,n+1}$ are uniquely obtained by
\begin{equation}
    \begin{bmatrix}
    \alpha_{i,2}\\
    \vdots\\
    \alpha_{i,n+1}
    \end{bmatrix}
    =\begin{bmatrix}
    \mathbf{r}_{2,0}-\mathbf{r}_{1,0}&\cdots&\mathbf{r}_{n+1,0}-\mathbf{r}_{1,0}
    \end{bmatrix}
    ^{-1}\mathbf{r}_{i,0}.
\end{equation}
Now, Eq. \eqref{pprrooff} can be written in the form of Eq. \eqref{affineeeeelead}, where $\alpha_{i,1}=1-\sum_{j=2}^{n+1}\alpha_{i,j}$ which in turn implies Eq. \eqref{alfasum}.

  

{\color{black}\textbf{Proof of Proposition \ref{PROP2222}:} 
Elements of matrix $\mathbf{U}_D=\left[U_{ij}\right]\in \mathbb{R}^{3\times 3}$, defined by \eqref{UDuseful}, are expressed as follows:
\[
\resizebox{0.99\hsize}{!}{%
$
\begin{split}
    U_{11}=&
\lambda_1\left(C_{{\beta_5}}^2C_{{\beta_6}}^2\right)+\lambda_2\left(S_{\beta_4}^2S_{\beta_5}^2C_{\beta_6}^2+C_{\beta_4}^2S_{\beta_6}^2-2S_{\beta_4}S_{\beta_5}C_{\beta_6}C_{\beta_4}S_{\beta_6}\right)\\
+&\lambda_3\left(C_{\beta_4}^2S_{\beta_5}^2C_{\beta_6}^2+S_{\beta_4}^2S_{\beta_6}^2-2S_{\beta_6}S_{\beta_4}C_{\beta_4}S_{\beta_5}C_{\beta_6}\right),
\end{split}
$
}
\]
\[
\resizebox{0.99\hsize}{!}{%
$
\begin{split}
    U_{12}=&
\lambda_1C_{\beta_6}C_{\beta_5}^2S_{\beta_6} - \lambda_3(C_{\beta_6}S_{\beta_4}^2S_{\beta_6} - C_{\beta_4}^2C_{\beta_6}S_{\beta_6}S_{\beta_5}^2) - \lambda_2(C_{\beta_4}^2C_{\beta_6}S_{\beta_6} \\
-& C_{\beta_6}S_{\beta_4}^2S_{\beta_6}S_{\beta_5}^2),
\end{split}
$
}
\]
\[
\resizebox{0.99\hsize}{!}{%
$
\begin{split}
    U_{13}=&
\lambda_3(C_{\beta_6}C_{\beta_5}S_{\beta_5}C_{\beta_4}^2 + C_{\beta_5}S_{\beta_4}S_{\beta_6}C_{\beta_4}) + \lambda_2(C_{\beta_6}C_{\beta_5}S_{\beta_4}^2S_{\beta_5} - C_{\beta_4}C_{\beta_5}S_{\beta_4}S_{\beta_6})\\ -& \lambda_1C_{\beta_6}C_{\beta_5}S_{\beta_5},
\end{split}
$
}
\]
\[
\resizebox{0.99\hsize}{!}{%
$
\begin{split}
    U_{22}=&
\lambda_2(C_{\beta_4}^2C_{\beta_6}^2 + S_{\beta_4}^2S_{\beta_6}^2S_{\beta_5}^2 + 2C_{\beta_4}C_{\beta_6}S_{\beta_4}S_{\beta_6}S_{\beta_5}) + \lambda_3(C_{\beta_6}^2S_{\beta_4}^2\\ +& C_{\beta_4}^2S_{\beta_6}^2S_{\beta_5}^2 - 2C_{\beta_4}C_{\beta_6}S_{\beta_4}S_{\beta_6}S_{\beta_5}) + \lambda_1C_{\beta_5}^2S_{\beta_6}^2,
\end{split}
$
}
\]
\[
    U_{33}=\lambda_3C_{\beta_4}^2C_{\beta_5}^2 + \lambda_2C_{\beta_5}^2S_{\beta_4}^2 + \lambda_1S_{\beta_5}^2,
\]
$U_{21}=U_{12}$, $U_{32}=U_{23}$, and $U_{31}=U_{13}$. If $\lambda_1=\lambda_2=\lambda_3=\lambda$, 
then,
\[
U_{ij}=\begin{cases}
    \lambda&i=j\\
    0&i\neq j
\end{cases}
.
\]
}
  \textbf{Proof of Theorem \ref{thmlambdamax}:} 
  Inter-agent collision between every two quadcopters is avoided, if 
\begin{equation}
\label{rawcollision}
    \forall i,j\in \mathcal{V},~i\neq j,\qquad \|\mathbf{r}_i(t)-\mathbf{r}_j(t)\|>2\epsilon.
\end{equation}
  We can write
  \[
\left(\mathbf{r}_{i}-\mathbf{r}_{j}\right)=\left(\mathbf{r}_{i,a}-\mathbf{r}_{j,a}\right)-\left(\mathbf{r}_{i,a}-\mathbf{r}_i\right)-\left(\mathbf{r}_j-\mathbf{r}_{j,a}\right).
\]
Therefore,
\begin{equation}
\label{prooffferror}
    \|\mathbf{r}_{i}-\mathbf{r}_{j}\|\geq \|\mathbf{r}_{i,a}-\mathbf{r}_{j,a}\|-\|\mathbf{r}_{i,a}-\mathbf{r}_i\|-\|\mathbf{r}_j-\mathbf{r}_{j,a}\|.
\end{equation}
Eq. \eqref{prooffferror} implies that  $\|\mathbf{r}_{i}-\mathbf{r}_{j}\|\leq 2\epsilon$, if 
$\|\mathbf{r}_{i,a}-\mathbf{r}_{i}\|\leq \delta$, $\|\mathbf{r}_{j,a}-\mathbf{r}_{j}\|\leq \delta$, and  $\|\mathbf{r}_{i,a}-\mathbf{r}_{j,a}\|\geq  2\left(\epsilon+\delta\right)$. Furthermore, $\|\mathbf{r}_{i,a}-\mathbf{r}_{j,a}\|\geq  2\left(\epsilon+\delta\right)$, if $\|\left(\mathbf{r}_{i,a}-\mathbf{r}_{j,a}\right)\cdot \hat{\mathbf{u}}_1\|\geq 2\left(\delta+\epsilon\right)$. Consequently, inter-agent collision avoidance between every two different quadcopters $i$ and $j$ is avoided, if 
\[
\|\left(\mathbf{r}_{i,a}-\mathbf{r}_{j,a}\right)\cdot \hat{\mathbf{u}}_1\|\geq 2\left(\delta+\epsilon\right).
\]

Note that
\[
\forall i,j\in \mathcal{V},~i\neq j,\qquad 
\lambda_{1}(t)=\dfrac{|\left(\mathbf{r}_{i,a}(t)-\mathbf{r}_{j,a}(t)\right)\cdot \hat{\mathbf{u}}_1(t)|}{|\left(\mathbf{r}_{i,0}(t_0)-\mathbf{r}_{j,0}(t_0)\right)\cdot \hat{\mathbf{u}}_1(t_0)|}
\]
It is ensured that no quadcopter leaver ball $\bar{\mathcal{S}}(t)$ at any time $t\in [t_0,t_f]$, if
{\color{black}
\[
\resizebox{0.99\hsize}{!}{%
$
\forall t\in [t_0,t_f],\qquad \max\limits_{i\in \mathcal{V}}\|\mathbf{r}_{i,a}(t)-\mathbf{d}\left(t\right) \|_2\leq r_{\mathrm{max}}-\delta-\epsilon.
$
}
\]
}
This implies that Eq. \eqref{lmax} assigns the upper-limit for eigenvalue $\lambda_1(t)$ at any time $t\in [t_0,t_f]$. Additionally, It is ensured that no two quadcopters collide, if
\[
\forall t\in [t_0,t_f],\qquad \min\limits_{i,j\in \mathcal{V},~i\neq j}\left|\left(\mathbf{r}_{i,a}(t)-\mathbf{r}_{j,a}(t)\right)\cdot \hat{\mathbf{u}}_1(t)\right|\geq 2\left(\delta+\epsilon\right).
\]
Consequently, Eq. \eqref{lmin} assigns the lower limit for $\lambda_1(t)$ at any time $t\in [t_0,t_f]$.

Because $\mathbf{\Omega}_0\subset \mathcal{S}(t_0)$ and  $\mathbf{\Omega}_0\subset \mathcal{S}(t_f)$. Therefore, $\lambda_{2,0}\leq \lambda_{\mathrm{max}}$, $\lambda_{2,f}\leq \lambda_{\mathrm{max}}$, $ \lambda_{3,0}\leq \lambda_{\mathrm{max}}$, $ \lambda_{3,f}\leq \lambda_{\mathrm{max}}$, and $\lambda_j(t)=\lambda_{j,0}\left(1-\gamma(t,T)\right)+\gamma(t,T)\lambda_{j,f}$ remains bounded:
\[
j=2,3,\qquad\lambda_j(t)\leq \lambda_{\mathrm{max}}
\]
at any time $t\in [t_0,t_f]$.

\textbf{Proof of Theorem \ref{th2}:}
If assumptions of Theorem \ref{th2} are all satisfied, matrix $\mathbf{G}\in \mathbf{R}^{\left(N-n-1\right)\times \left(N-n-1\right)}$ is non-negative a there exists a directed path between every leader and every follower. By provoking Perron-Frobenius Theorem, it is concluded that the spectral radius of matrix $\mathbf{G}$, denoted by $\rho\left(\mathbf{G}\right)$, is less than $1$ and eigenvalues of matrix $-\mathbf{I}+\mathbf{G}$ are all placed on the left-hand of the $s$-plane inside a disk of radius $\rho\left(\mathbf{G}\right)$ centered at $-1+0\mathrm{j}$. Therefore, matrices $-\mathbf{I}+\mathbf{G}$ and $\mathbf{L}=-\mathbf{I}+\mathbf{W}$ are Hurwitz.

\textbf{Proof of Theorem \ref{Lem1}:}
Let 
$
\mathbf{R}_{L,0}=\mathrm{vec}\left(
\begin{bmatrix}
\mathbf{r}_{1,0}&\cdots&\mathbf{r}_{n+1,0}
\end{bmatrix}^T\right)\in \mathbb{R}^{3\left(n+1\right)\times 1}
$ and
$
\mathbf{R}_{F,0}=\mathrm{vec}\left(
\begin{bmatrix}
\mathbf{r}_{n+2,0}&\cdots&\mathbf{r}_{N,0}
\end{bmatrix}^T\right)\in \mathbb{R}^{3\left(N-n-1\right)\times 1},
$
define initial position components of leaders and  followers, respectively, where $\mathrm{vec}\left(\cdot\right)$ is the matrix vectorization operator. If assumptions of Lemma \ref{Lem1} are satisfied, quadcopters' initial positions satisfy the following relation:
\[
\mathbf{L}
\begin{bmatrix}
\mathbf{R}_{L,0}\\
\mathbf{R}_{F,0}
\end{bmatrix}
=
\left[
\begin{array}{cc}
    -\mathbf{I} &\mathbf{0}  \\
   \mathbf{F}&\left(-\mathbf{I}+\mathbf{G}\right)
\end{array}
\right]
\begin{bmatrix}
\mathbf{R}_{L,0}\\
\mathbf{R}_{F,0}
\end{bmatrix}
=
\begin{bmatrix}
-\mathbf{R}_{L,0}\\
\mathbf{0}
\end{bmatrix}
.
\]
Thus, 
\[
\begin{bmatrix}
\mathbf{R}_{L,0}\\
\mathbf{R}_{F,0}
\end{bmatrix}
=\begin{bmatrix}
\mathbf{I}\\
-\left(-\mathbf{I}+\mathbf{G}\right)^{-1}\mathbf{F}
\end{bmatrix}
\mathbf{R}_{L,0}. 
\]
Because leaders form an $n$-D simplex at initial time $t_0$, positions of every quadcopter $i$ can be uniquely expressed as a linear combination of leaders' initial positions  using relation \eqref{affineeeeelead}. Therfore,
\[
\begin{bmatrix}
\mathbf{R}_{L,0}\\
\mathbf{R}_{F,0}
\end{bmatrix}
=\mathbf{H}
\mathbf{R}_{L,0}
\]
which in turn implies correctness of Eq. \eqref{secondwl}.

Now, we can write
\begin{equation}
    \mathbf{Y}_d-\mathbf{Y}=\left(\mathbf{I}_3\otimes \mathbf{L}\right)\mathbf{Y}+\left(\mathbf{I}_3\otimes \mathbf{L}_0\right)\mathbf{R}_L(t)
\end{equation}
On the other hand,
\begin{equation}
    \left(\mathbf{I}_3\otimes\mathbf{L}\right)\left(\mathbf{Y}-\mathbf{Y}_a\right)=
    \mathbf{Y}_d-\mathbf{Y}.
\end{equation}
Therefore, Eq. \eqref{y-ya} is proven.
\\
{\color{black}
\textbf{Proof of Theorem  \ref{theorem9999}:} Given definition of $\gamma\left(t,T_l\right)$ in  \eqref{poly1}, $\dot{\gamma}\left(t,T_l\right)$, $\ddot{\gamma}\left(t,T_l\right)$, and $\dddot{\gamma}\left(t,T_l\right)$ are decreasing with respect to $T_l$. {\color{black}For a given initial time $t_0$, $T_l$, defined by \eqref{TL}, is increased if $t_f$ is increased.} Also, $\dot{\gamma}\left(t,T_l\right),\ddot{\gamma}\left(t,T_l\right),\dddot{\gamma}\left(t,T_l\right)\rightarrow 0$, and $\mathbf{V}_{\mathrm{MQS}}\rightarrow \mathbf{0}$, if $T_f\rightarrow \infty$. Therefore, there exists a sufficiently-large final time $\tilde{t}_f=t_0+\tilde{T}_1+\cdots+\tilde{T}_{n_\tau-1}$ such that the response of {\color{black}zero-initial-state dynamics, given by
\[
\begin{bmatrix}
\mathbf{E}(t)\\
\dot{\mathbf{E}}(t)\\
\ddot{\mathbf{E}}(t)\\
\dddot{\mathbf{E}}(t)\\
\end{bmatrix}
=\mathrm{e}^{\mathbf{A}_{\mathrm{MQS}}{\left(t-t_0\right)}}
    {\color{black}\cancelto{\mathbf{0}}{
\begin{bmatrix}
\mathbf{E}(t_0)\\
\dot{\mathbf{E}}(t_0)\\
\ddot{\mathbf{E}}(t_0)\\
\dddot{\mathbf{E}}(t_0)\\
\end{bmatrix}}}+
\int_{t_0}^t\mathrm{e}^{\mathbf{A}_{\mathrm{SYS}}\left(t-\tau\right)}{\mathbf{V}_{\mathrm{MQS}}\left(\tau\right)}\mathrm{d}\tau,
\]
ensures that $\|\mathbf{r}_i(t)-\mathbf{r}_{i,a}(t)\|<(1-\varrho) \delta$} for every quadcopter $i\in \mathcal{V}$. This also ensures that safety condition \eqref{eqreferror12} is satisfied, if the zero dynamics of the error dynamics \eqref{errorlinearized} ensures that $\|\mathbf{r}_i(t)-\mathbf{r}_{i,a}(t)\|<\varrho \delta$ at any time $t\in [t_0,t_f]$.
}

\vspace{-.3cm}
\section{{\color{black}Rotational Kinematics and Dynamics of a Quacopter}}\label{Quadcopter Angular Velocity and Acceleration}
We use 3-2-1 standard to determine orientation of quadcopter $i$ at discrete time $k$. Given roll angle $\phi_i(t)$, pitch angle $\theta_i$, and yaw angle $\psi_i$ and the base vectors of the inertial coordinate system ($\hat{\mathbf{e}}_1$, $\hat{\mathbf{e}}_2$, and $\hat{\mathbf{e}}_3$), angular velocity of quadcopter $i\in \mathcal{V}$ is given by
\begin{equation}
\label{AngularVelocity}
    {\color{black}\bf{\omega}}_i=\dot{\psi}_i\hat{\mathbf{k}}_{1,i}+\dot{\theta}_i\hat{\mathbf{j}}_{2,i}+\dot{\phi}_i\hat{\mathbf{i}}_{b,i},
\end{equation}
where
\begin{subequations}
\begin{equation}
    \begin{bmatrix}
\hat{\mathbf{i}}_{1,i}\\
\hat{\mathbf{j}}_{1,i}\\
\hat{\mathbf{k}}_{1,i}
\end{bmatrix}
=\mathbf{R}\left(0,0,\psi_{i}\right)
\begin{bmatrix}
\hat{\mathbf{e}}_{1}\\
\hat{\mathbf{e}}_{2}\\
\hat{\mathbf{e}}_{3}
\end{bmatrix}
=\begin{bmatrix}
\cos\psi_i&\sin\psi_i&0\\
-\sin\psi_i&\cos\psi_i&0\\
0&0&1\\
\end{bmatrix}
\begin{bmatrix}
\hat{\mathbf{e}}_{1}\\
\hat{\mathbf{e}}_{2}\\
\hat{\mathbf{e}}_{3}
\end{bmatrix}
,
\end{equation}
\begin{equation}
    \begin{bmatrix}
\hat{\mathbf{i}}_{2,i}\\
\hat{\mathbf{j}}_{2,i}\\
\hat{\mathbf{k}}_{2,i}
\end{bmatrix}
=\mathbf{R}\left(0,\theta_i,0\right)
\begin{bmatrix}
\hat{\mathbf{i}}_{1,i}\\
\hat{\mathbf{j}}_{1,i}\\
\hat{\mathbf{k}}_{1,i}
\end{bmatrix}
=\begin{bmatrix}
\cos\theta_i&0&-\sin\theta_i\\
0&1&0\\
\sin\theta_i&0&\cos\theta_i\\
\end{bmatrix}
\begin{bmatrix}
\hat{\mathbf{i}}_{1,i}\\
\hat{\mathbf{j}}_{1,i}\\
\hat{\mathbf{k}}_{1,i}
\end{bmatrix}
,
\end{equation}
\begin{equation}
    \begin{bmatrix}
\hat{\mathbf{i}}_{b,i}\\
\hat{\mathbf{j}}_{b,i}\\
\hat{\mathbf{k}}_{b,i}
\end{bmatrix}
=\mathbf{R}\left(\phi_{i},0,0\right)
\begin{bmatrix}
\hat{\mathbf{i}}_{2,i}\\
\hat{\mathbf{j}}_{2,i}\\
\hat{\mathbf{k}}_{2,i}
\end{bmatrix}
=\begin{bmatrix}
1&0&0\\
0&\cos\phi_i&\sin\phi_i\\
0&-\sin\phi_i&\cos\phi_i
\end{bmatrix}
\begin{bmatrix}
\hat{\mathbf{i}}_{2,i}\\
\hat{\mathbf{j}}_{2,i}\\
\hat{\mathbf{k}}_{2,i}
\end{bmatrix}
.
\end{equation}
\end{subequations}
Substituting $\hat{\mathbf{e}}_1=\left[1~0~0\right]^T$, $\hat{\mathbf{e}}_2=\left[0~1~0\right]^T$, $\hat{\mathbf{e}}_3=\left[0~0~1\right]^T$, $\hat{\mathbf{i}}_{1,i}$, $\hat{\mathbf{k}}_{1,i}$, $\hat{\mathbf{j}}_{2,i}$, and $\hat{\mathbf{i}}_{b,i}$ into Eq. \eqref{AngularVelocity}, ${\color{black}\bf{\omega}}_i=\left[\omega_{x,i}~\omega_{y,i}~\omega_{z,i}\right]^T$ is related by $\dot{\phi}_i$, $\dot{\theta}_i$, and $\dot{\psi}_i$ by
\begin{equation}
    \begin{bmatrix}
    \omega_{x,i}&
    \omega_{y,i}&
    \omega_{z,i}
    \end{bmatrix}^T
    =\mathbf{\Gamma}_i\left(\phi_i,\theta_i,\psi_i\right)
    \begin{bmatrix}
        \dot{\phi}_{i}&
        \dot{\theta}_{i}&
        \dot{\psi}_{i}
    \end{bmatrix}
    ^T
    ,
\end{equation}
where
\begin{equation}
\label{Eq55}
\mathbf{\Gamma}_i\left(\phi_i,\theta_i,\psi_i\right)=
    \begin{bmatrix}
    1&0&-\sin\theta_i\\
    0&\cos\phi_i&\cos\theta_i\sin\phi_i\\
    0&-\sin\phi_i&\cos\phi_i\cos\theta_i
    \end{bmatrix}
    .
\end{equation}

Angular acceleration of quadcopter $i\in \mathcal{V}$ is obtained by taking {\color{black}the} time derivative {\color{black}of} the angular velocity vector ${\color{black}\bf{\omega}}_i$:
\begin{equation}
\label{dwegai1}
\begin{split}
    \dot{{\color{black}\bf{\omega}}}_i=
    \tilde{\mathbf{B}}_{1,i}\begin{bmatrix}
    \ddot{\phi}_i&\ddot{\theta}_i&\ddot{\psi}_i
    \end{bmatrix}^T+\tilde{\mathbf{B}}_{2,i}.
\end{split}
\end{equation}
where
\begin{subequations}
\begin{equation}
    \tilde{\mathbf{B}}_{1,i}=\begin{bmatrix}
    \hat{\mathbf{i}}_{b,i}&\hat{\mathbf{j}}_{2,i}&\hat{\mathbf{k}}_{1,i}
    \end{bmatrix}
\end{equation}
\begin{equation}
   \tilde{\mathbf{B}}_{2,i}=\dot{\theta}_i\dot{\psi}_i\left(\hat{\mathbf{k}}_{1,i}\times \hat{\mathbf{j}}_{1,i} \right)+\dot{\phi}_{i}\left(\dot{\psi}_i\hat{\mathbf{k}}_{1,i}+\dot{\theta}_i\hat{\mathbf{j}}_{2,i}\right)\times \hat{\mathbf{i}}_{2,i}
\end{equation}
\end{subequations}
{\color{black}Note that ``$\times $'' is the cross product symbol. On the other hand, the rotational dynamics of quadcopter $i$ is given by
\begin{equation}
\label{dwegai2}
    \dot{\bf{\omega}}_i=
    \mathbf{J}_{i}^{-1}\left(\bf{\omega}_i\times \left(\mathbf{J}_i\bf{\omega}_i\right)+\begin{bmatrix}
    \tilde{u}_{2,i}&\tilde{u}_{3,i}&\tilde{u}_{4,i}
    \end{bmatrix}^T\right)
\end{equation}
where $\tilde{u}_{2,i}=\tau_{\phi,i}$, $\tilde{u}_{3,i}=\tau_{\theta,i}$, $\tilde{u}_{4,i}\tau_{\psi,i}$ (See Eq. \eqref{extendednonlineardynamicsquadcopter}). 
By equating the right-hand sides of Eqs. \eqref{dwegai1} and \eqref{dwegai2}, we can write
\begin{equation}
\label{88badeq}
    \begin{bmatrix}
    \tilde{u}_{2,i}&\tilde{u}_{3,i}&\tilde{u}_{4,i}
    \end{bmatrix}^T=\mathbf{B}_{1,i}\begin{bmatrix}
    \ddot{\phi}_i&\ddot{\theta}_i&\ddot{\psi}_i
    \end{bmatrix}^T+\mathbf{B}_{2,i},
\end{equation}
where
\begin{subequations}
\begin{equation}
    \mathbf{B}_{1,i}=\mathbf{J}_i\tilde{\mathbf{B}}_{1,i},
\end{equation}
\begin{equation}
    \mathbf{B}_{2,i}=\mathbf{J}_i\tilde{\mathbf{B}}_{2,i}-\bf{\omega}_i\times \left(\mathbf{J}_i\bf{\omega}_i\right).
\end{equation}
\end{subequations}
}

\section{Time Derivatives of the Quadcopter Thrust Force}\label{Time Derivatives of the Quadcopter Thrust Force}
{\color{black}Let 
\begin{equation}
    \mathbf{P}_i=p_i\hat{\mathbf{k}}_{b,i}-m_ig\hat{\mathbf{e}}_1
\end{equation}
be the external force executed on quadcopter $i$.
}Taking time derivatives from $\mathbf{P}_i$, we obtain the following relations:{\color{black}
\begin{subequations}
\begin{equation}
    \dot{\mathbf{P}}_i=\dot{p}_i\hat{\mathbf{k}}_{b,i}+p_i{\color{black}\bf{\omega}}_i\times \hat{\mathbf{k}}_{b,i},
\end{equation}
\begin{equation}
\label{doubledotFii}
\ddot{\mathbf{P}}_i
=\mathbf{O}_{1,i}\mathbf{\Xi}_i+\mathbf{O}_{2,i},
\end{equation}
\end{subequations}
where $\mathbf{\Xi}_i=\begin{bmatrix}
    \ddot{p}_i&
    \ddot{\phi}_i&
    \ddot{\theta}_i
  &\ddot{\psi}_i
    \end{bmatrix}^T$,
\begin{subequations}
\label{MiNi}
\begin{equation}
\mathbf{O}_{1,i}=\left[
\begin{array}{cccc}
    \hat{\mathbf{k}}_{b,i}&-p_i\hat{\mathbf{j}}_{b,i}&p_i\left(\hat{\mathbf{j}}_{2,i}\times \hat{\mathbf{k}}_{b,i}\right)& p_i\hat{\mathbf{k}}_{1,i}\times\hat{\mathbf{k}}_{b,i} 
\end{array}
\right] \in \mathbb{R}^{3\times 4},
\end{equation}
\begin{equation}
\begin{split}
    \mathbf{O}_{2,i}=&p_i\left[-\dot{\phi}_i\dot{\theta}_i\left(\hat{\mathbf{k}}_{2,i}\times \hat{\mathbf{k}}_{b,i}\right)+{\color{black}\bf{\omega}}_i\times \left({\color{black}\bf{\omega}}_i\times \hat{\mathbf{k}}_{b,i}\right)\right]+2\dot{p}_i{\color{black}\bf{\omega}}_i\times \hat{\mathbf{k}}_{b,i}.
\end{split}
\end{equation}
\end{subequations}
Per Eq. \eqref{88badeq}, we can write 
\begin{equation}
\label{91bbbbbbbbbbbbbb}
    \mathbf{\Xi}_i=\mathbf{O}_{3,i}\tilde{\mathbf{u}}_i+\mathbf{O}_{4,i},
\end{equation}
where $\tilde{\mathbf{u}}_i=\begin{bmatrix}
\tilde{u}_{1,i}&\cdots&\tilde{u}_{4,i}
\end{bmatrix}^T=\begin{bmatrix}
\ddot{p}_i&\tau_{\phi,i}&\tau_{\theta,i}&\tau_{\psi,i}
\end{bmatrix}^T$,
\[
\mathbf{O}_{3,i}=\left[
\begin{array}{c|c}
1&\mathbf{0}_{1\times 3}\\
\hline
\mathbf{0}_{3\times 1}&\mathbf{B}_{1,i}^{-1}
\end{array}
\right]
,
\]
\[
\mathbf{O}_{4,i}=\left[
\begin{array}{c}
0\\
\hline
-\mathbf{B}_{1,i}^{-1}\mathbf{B}_{2,i}
\end{array}
\right]
.
\]
By substituting  \eqref{91bbbbbbbbbbbbbb}, Eq. \eqref{doubledotFii} is converted to
\begin{equation}
\label{doubledotFi}
\ddot{\mathbf{P}}_i
=\mathbf{O}_{1,i}\mathbf{O}_{3,i}\tilde{\mathbf{u}}_i+\mathbf{O}_{1,i}\mathbf{O}_{4,i}+\mathbf{O}_{2,i}.
\end{equation}
Note that  $\mathbf{s}_i={1\over m_i}\ddot{\mathbf{P}}_i$ where $\mathbf{s}_i$ is the input vector of the external dynamics of quadcopter $i$ (see Section \ref{MAS Collective Dynamics1}). 

}
\begin{IEEEbiography}[{\includegraphics[width=1in,height=1.25in,clip,keepaspectratio]{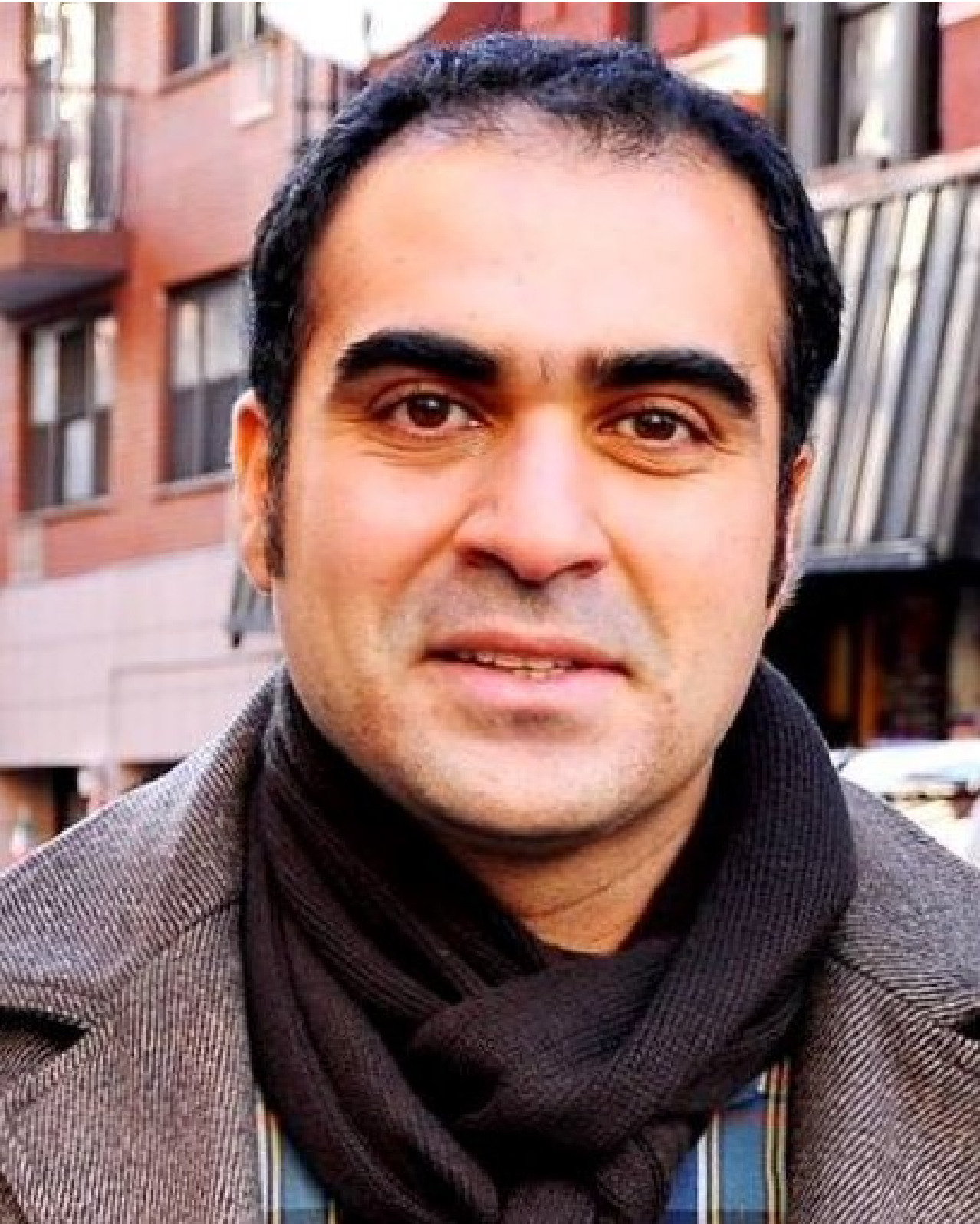}}]
{\textbf{Hossein Rastgoftar}} an Assistant Professor at Villanova University and an Adjunct Assistant Professor at the University of Michigan. He was an Assistant Research Scientist in the Aerospace Engineering Department from 2017 to 2020. Prior to that he was a postdoctoral researcher at the University of Michigan from 2015 to 2017. He received the B.Sc. degree in mechanical engineering-thermo-fluids from Shiraz University, Shiraz, Iran, the M.S. degrees in mechanical systems and solid mechanics from Shiraz University and the University of Central Florida, Orlando, FL, USA, and the Ph.D. degree in mechanical engineering from Drexel University, Philadelphia, in 2015. 
\end{IEEEbiography}
\begin{IEEEbiography}[{\includegraphics[width=1in,height=1.2in,clip,keepaspectratio]{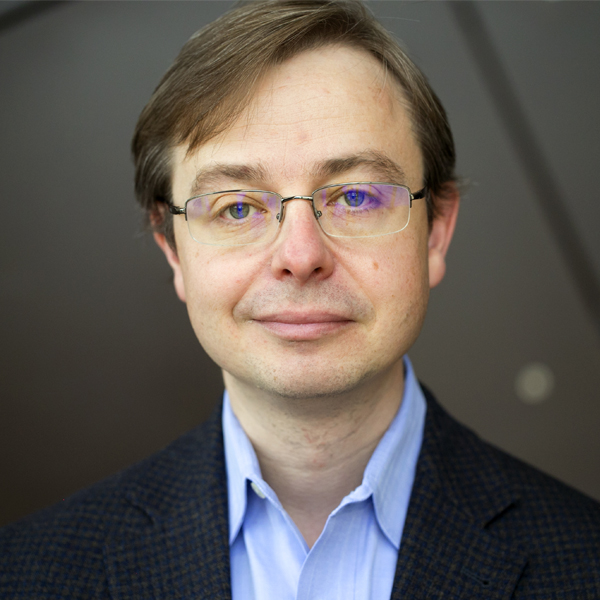}}]
{\textbf{Ilya V. Kolmanovsky}} received M.S. and Ph.D. degrees in aerospace engineering and the M.A. degree in mathematics from the University of Michigan, Ann Arbor, MI, USA, in 1993, 1995, and 1995, respectively.  Between 1995 and 2009, he was with Ford Research and Advanced Engineering, Dearborn, MI, USA. He is currently a Full Professor with the Department of Aerospace Engineering, University of Michigan. His  research interests include control theory for systems with state and control constraints, and control applications to aerospace and automotive systems. Dr. Kolmanovsky was a recipient of the Donald P. Eckman Award of American Automatic Control Council and two IEEE Transactions on Control Systems Technology Outstanding Paper Awards. Dr. Kolmanovsky is an IEEE Fellow and {\color{black}AIAA Associate Fellow}.
\end{IEEEbiography}

\end{document}